\documentclass{ieeeaccess}
\usepackage{cite}
\usepackage{tikz}
\NewSpotColorSpace{PANTONE}
\AddSpotColor{PANTONE} {PANTONE3015C} {PANTONE\SpotSpace 3015\SpotSpace C} {1 0.3 0 0.2}
\SetPageColorSpace{PANTONE}
\usepackage{amsmath,amssymb,amsfonts}
\usepackage{bm}
\usepackage{algorithmic}
\usepackage{algorithm}
\usepackage{array}
\usepackage[caption=false,font=normalsize,labelfont=sf,textfont=sf]{subfig}
\usepackage{textcomp}
\usepackage{stfloats}
\usepackage{hyperref}
\usepackage{xurl}
\hypersetup{breaklinks=true}
\usepackage{verbatim}
\usepackage{graphicx}
\usepackage{booktabs}
\usepackage{multirow}
\usepackage{colortbl}
\usepackage{tcolorbox}
\usepackage{tabularx}
\usepackage{makecell}
\usepackage{standalone}
\usepackage{tcolorbox}
\usepackage{enumitem}
\usepackage{listings}
\usepackage{microtype}
\usepackage{multibib}
\usepackage{comment}
\newcites{r}{Reviewed Literature}

\makeatletter
\AtBeginDocument{\DeclareMathVersion{bold}
\SetSymbolFont{operators}{bold}{T1}{times}{b}{n}
\SetSymbolFont{NewLetters}{bold}{T1}{times}{b}{it}
\SetMathAlphabet{\mathrm}{bold}{T1}{times}{b}{n}
\SetMathAlphabet{\mathit}{bold}{T1}{times}{b}{it}
\SetMathAlphabet{\mathbf}{bold}{T1}{times}{b}{n}
\SetMathAlphabet{\mathtt}{bold}{OT1}{pcr}{b}{n}
\SetSymbolFont{symbols}{bold}{OMS}{cmsy}{b}{n}
\renewcommand\boldmath{\@nomath\boldmath\mathversion{bold}}}
\makeatother

\newcommand{\imagedir}{./images/}

\renewcommand{\thevol}{X}
\renewcommand{\theyear}{X}

\begin{document}

\usetikzlibrary{positioning, arrows, arrows.meta, calc, automata, shapes, shadows, 
fit}

\history{Date of current version 24 June 2024.}
\doi{}

\title{Migrating Software Systems Towards Post-Quantum Cryptography -- A Systematic
Literature Review}

\author{
  \uppercase{Christian Näther}\authorrefmark{1}, 
  \uppercase{Daniel Herzinger}\authorrefmark{2}, 
  \uppercase{Stefan-Lukas Gazdag}\authorrefmark{2}, 
  \uppercase{Jan-Philipp Steghöfer}\authorrefmark{1}, \IEEEmembership{Member, IEEE},
  \uppercase{Simon Daum}\authorrefmark{2}, 
  AND \uppercase{Daniel Loebenberger}\authorrefmark{3}
}

\address[1]{Xitaso GmbH, Augsburg, 86153 Germany}
\address[2]{genua GmbH, Kirchheim, 85551 Germany}
\address[3]{Fraunhofer AISEC, Garching, 85748 Germany}

\tfootnote{The work was funded by the German Federal Ministry of Education and Research.}

\markboth
{Näther \headeretal: Preparation of Papers for IEEE ACCESS}
{Näther \headeretal: Preparation of Papers for IEEE ACCESS}

\corresp{Corresponding author: Christian Näther (e-mail: christian.naether@xitaso.com)}

\begin{abstract}
  Networks such as the Internet are essential for our connected world.
  Quantum computing threatens its fundamental security mechanisms. Therefore, a migration to post-quantum-cryptography (PQC) is necessary for networks and their components.
  Currently, there is little knowledge on how such migrations should be structured and implemented in practice. Our systematic literature review addresses migration approaches for IP networks towards PQC. It surveys papers about the migration process and exemplary real-world software system migrations. On the process side, we found that terminology, migration steps, and roles are not defined precisely or consistently across the literature. Still, we identified four major phases and appropriate substeps which we matched with also emerging archetypes of roles. In terms of real-world migrations, we see that reports used several different PQC implementations and hybrid solutions for migrations of systems belonging to a wide range of system types. Across all papers we noticed three major challenges for adopters: missing experience of PQC and a high realization effort, concerns about the security of the upcoming system, and finally, high complexity. Our findings indicate that recent standardization efforts already push quantum-safe networking forward. However, the literature is still not in consensus about definitions and best practices. Implementations are mostly experimental and not necessarily practical, leading to an overall chaotic situation. To better grasp this fast moving field of (applied) research, our systematic literature review provides a comprehensive overview of its current state and serves as a starting point for delving into the matter of PQC migration.
\end{abstract}


\titlepgskip=-21pt
\maketitle

\let\thefootnote\relax\footnotetext{This work has been submitted to the IEEE for possible publication. Copyright may be transferred without notice, after which this version may no longer be accessible.}

\section{Introduction}
\label{seq:introduction}
The awareness of potential future attacks based on quantum computers is being
raised continuously~\cite{status_qc_bsi}. As the disruptive technology of quantum
computing may cause a cryptographic apocalypse by overthrowing or weakening secure
building blocks of our digital communication, the call for quantum-safe solutions is
rising as well (compare, e.g., \cite{bsigestalten,white_house_memorandum}). Despite
the enormous efforts around the globe to provide such future-proof alternatives, the past
catches up with us. Networking\,---\,meaning the Internet as well as private IP
networks connected to the Internet\,---\,is based on a wide range of very old to
brand new technologies. Updating this ever-growing infrastructure is a difficult task
on its own and suitable frameworks usually encounter a major question which is not
sufficiently addressed in research and practice:
\emph{How to migrate to quantum-safe mechanisms in such a complex ecosystem given
the diverse requirements and constraints by protocols and (non-standard)
implementations?}

Practical knowledge about this
transition or migration\,---\,especially on a broader scale\,---\,is
limited (e.g., \cite{google-experiment,cloudflare-tls}). Public institutions,
companies, and organizations are starting with the first
relevant steps such as creating cryptographic inventories and undertaking
risk assessment. While there are various
ways to perform these tasks, especially the actual transition can only be described
in either a very abstract way or by deep-diving into very specific use-cases. Over
time this will likely converge, but for now no comprehensive account of a practical
transition can be given. Not only people new to the topic are overwhelmed by the
sheer number of academic papers, reports, handbooks, and
other documents available. During an initial literature assessment several hundred publications emerged. Though the work behind these
is important and relevant, deducing a concrete guideline for the transition is
difficult.
By means of a Systematic Literature Review (SLR) we try to identify the most relevant publications available.

\subsection*{Related Secondary Studies}
\label{s:introduction:related_secondary_studies}
As the research related to PQC and in particular migration to PQC is still
comparatively new, only a small set of secondary studies are available in
that field.

A 2020 study in this area was conducted by Lohachab et al.~\cite{iot_survey}. There, the
authors examined several aspects of PQC in IoT networks, including migration. The
authors' broader approach means, however, that there is no in-depth look at migration
and that it is entirely confined to the IoT context.

A year later, a secondary study by Wiesmaier et al.~\cite{on_migration}
provides a comprehensive survey of the literature about migration
and cryptographic agility up to 2021. The fast moving topic of PQC research and the fact
that the manuscript is a preprint that was not yet peer-reviewed as to our
knowledge, demands a more recent look at the topic.

Partly the same team of authors as from~\cite{on_migration} also published
Alnahawi et al.~\cite{on_ca} in 2023. That study is similar to Wiesmaier et al.~\cite{on_migration}
but focuses solely on the aspect of cryptographic agility and not PQC migration.

In a 2022 study by Giron et al.~\cite{hybrid_mapping}, the authors examine the aspects of hybrid PQC key exchanges. Those are widely seen as an
important building block for network security during the migration towards PQC
architectures.

Those secondary studies give insights either into topics adjacent to PQC migration or
into subtopics of it. Neither of them addresses the PQC migration systematically and
holistically.

Therefore, our study identifies and characterizes the state of the art regarding the
migration of existing software systems towards quantum-resistance
by means of a Systematic Literature Review (SLR). We identify
the most relevant publications available, extract concordant information on the
migration, and reduce bias by a reproducible procedure.
Thus, our work benefits the PQC community and potential implementers as
an overview and starting point into the current state of this field of research.

\section{Research Methodology}
\label{s:methodology}
In the following, we describe our research methodology.
Section~\ref{s:methodology:process} explains the general process that was used to
conduct the systematic literature review and its individual steps.
The first step was to identify related secondary studies, which we have already
described in Section~\ref{s:introduction:related_secondary_studies}.
Afterwards, we outline the definition of the research questions
in Section~\ref{s:methodology:research_questions} and the outline of a
suitable protocol to identify relevant literature in
Section~\ref{s:methodology:protocol_definition}.
Section~\ref{s:methodology:reference_set} describes how we compiled the reference set
of relevant papers used in quality control of the search protocol.
Section~\ref{s:methodology:search_and_selection} describes the search and
selection process for relevant literature. Section~\ref{s:methodology:data}
outlines data extraction and synthesis.

\subsection{Overview of the Research Process}
\label{s:methodology:process}
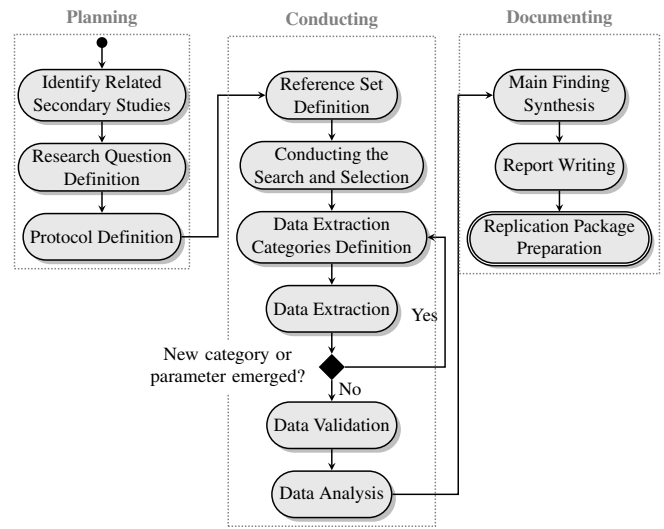
\begin{figure}[tb]
	\centering
  \begin{tikzpicture} [
  sta/.style={state, align=center, drop shadow},
  angle/.style={shading angle=45},
  box/.style={draw=black, line width=1, minimum width=200, minimum
		height=100, anchor=west, angle, align=center},
  component/.style={rectangle, draw=gray, fill=gray,
                    line width=1.5},
  blue_box/.style={top color=blue!35, bottom color=blue, box},
  red_box/.style={top color=red!35, bottom color=red, box},
  green_box/.style={top color=green!35, bottom color=green, box},
  gray_box/.style={top color=gray!35, bottom color=gray, box},
  arrow/.style={draw=black, line width=3, stealth},
	->,
	thick,
	>=stealth,
	auto,
	every initial by arrow/.style={*->},
	initial distance=0.6cm,
	initial text={},
	node distance=0.4cm
]
  \tikzstyle{every state}=[fill={rgb:black,1;white,10}, rounded
	rectangle]

	\node[sta, initial above] (2nd_studies) {\normalsize Identify Related\\Secondary
	  Studies};
	\node[sta, below=of 2nd_studies] (rq_def) {Research Question\\Definition};
	\node[sta, below=of rq_def] (prot_def) {Protocol Definition};
	\path (2nd_studies) edge (rq_def);
	\path (rq_def) edge (prot_def);
	\coordinate (up) at ([yshift=0.5cm]2nd_studies.north);
	\node[draw=gray, densely dotted, fit=(up)(prot_def), inner sep=0.15cm,
		label={[gray]:\textbf{Planning}}] (planning) {};

	\node[sta, right=1.5 cm of 2nd_studies] (ref_set) {Reference Set\\Definition};
	\node[sta, below=of ref_set] (search_sel) {Conducting the\\Search and Selection};
	\node[sta, below=of search_sel] (ext_def) {Data Extraction\\Categories Definition};
	\node[sta, below=of ext_def] (ext) {Data Extraction};
  \node[diamond, draw, fill, below=of ext] (decide) {};
	\draw (prot_def) -- ++(2.2cm, 0) |- (ref_set);
	\path (ref_set) edge (search_sel);
	\path (search_sel) edge (ext_def);
	\path (ext_def) edge (ext);
	\path (ext) edge (decide);

	\node[sta, below=of decide] (valid) {Data Validation};
	\node[sta, below=of valid] (anal) {Data Analysis};
	\draw (decide) -- (valid) node[pos=0.4] {No};
	\draw (decide) -- ++(2.15cm, 0) |- (ext_def) node[pos=0.2, left] {Yes};
	\path (valid) edge (anal);
	\coordinate (up1) at ([yshift=0.5cm]ref_set.north);
	\node[draw=gray, densely dotted, fit=(up1)(ext_def)(anal), inner sep=0.15cm,
		label={[gray]:\textbf{Conducting}}] (conducting) {};
	\node[left=-0 cm of decide, text width=3.2cm, align=center, fill=white] (condition)
	  {New category or\\parameter emerged?};

	\node[sta, right=1.8 cm of ref_set] (synth) {Main Finding\\Synthesis};
	\node[sta, below=of synth] (report) {Report Writing};
	\node[sta, below=of report, accepting] (repl) {Replication Package\\Preparation};
	\draw (anal) -- ++(2.4cm, 0) |- (synth);
	\path (synth) edge (report);
	\path (report) edge (repl);
	\coordinate (up2) at ([yshift=0.5cm]synth.north);
	\node[draw=gray, densely dotted, fit=(up2)(repl), inner sep=0.15cm,
		label={[gray]:\textbf{Documenting}}] (documenting) {};

\end{tikzpicture}
      \caption{Procedure of our search and selection process.}
  \label{fig:search_and_selection_procedure}
\end{figure}

Our research process is based on the PRISMA 2020 statement~\cite{prisma_2020}
and Kitchenham et al.~\cite{kitchenham_2013},
which serves as a guideline for transparent and uniform systematic reviews. We
systematically gathered and analyzed relevant literature, adhering to a three-step
approach: planning, conducting, and documenting the review.
This methodology has emerged as well-es\-tab\-lished in various relevant
publications~\cite{kitchenham_2013,petersen_slr_guidelines_2015,wohlin_experimentation_2012}.
Figure~\ref{fig:search_and_selection_procedure} and the subsequent subsections illustrate
how we implement the procedure using the three phases and their corresponding steps.

\subsubsection{Planning}
The planning phase provides the necessary foundation for the successful execution of
the subsequent phases. To avoid duplication and to create a working basis,
related secondary studies in a similar scope and context
were identified, as shown in Section~\ref{s:introduction:related_secondary_studies}.
Since these studies lack concrete and holistic guidance for PQC migration, we defined research questions (see
Section~\ref{s:methodology:research_questions}) and designed a research protocol to answer these questions
(see Section~\ref{s:methodology:protocol_definition}). More specifically, the research protocol
provides a comprehensive account of the steps we undertook during the different
phases. The protocol is based on Moher et al.~\cite{prisma-p_2015}, which contains the preferred
reporting items for systematic literature protocols.

\subsubsection{Conducting}
The Conducting Phase involves the identification, extraction, and analysis of
relevant data. We start by defining a reference set which
contains a number of manually selected papers that serve as a guideline for the
subsequent steps (see Section~\ref{s:methodology:reference_set}). A fixed number of
databases and search engines were then used to
capture the most relevant articles. These were matched against extraction criteria,
explicitly designed to specify the content as well as additional criteria, such as that
the sources we used had to be written in English.
Once the extraction criteria have been defined (see
Section~\ref{s:methodology:data_extraction_categories}), the next step is to extract
the data according to our quality assurance plan.
After the actual extraction, the data is then analyzed.

\subsubsection{Documenting}
The documenting phase is the third and final phase. The first step is to synthesize
the previously analyzed data by comparing the findings and preparing a condensed
shared truth. After writing the report, the final step is the preparation and
provision of a replication package. The aim of this package is to make our work
available to other researchers so that they can verify and reuse the
data.

\subsection{Research Questions}
\label{s:methodology:research_questions}
The three central research questions of this study are a refinement
of the overarching question we posed in Section~\ref{seq:introduction}.

\begin{description}
\item[RQ1:] \emph{What are the identified steps for the migration of software
systems towards post-quantum cryptography?}
\begin{quote}
To ensure that a PQC migration can be carried out, a corresponding process must be
available. We aim to provide a classification of the fundamental phases that need
to be done for a successful post-quantum migration of a software system by answering
RQ1. We also intend to refine some of the phases into concrete steps for an even
more precise and practically applicable procedure. Apart from the migration phases
and their steps, the roles that take on responsibility for those tasks will also
be covered within RQ1. This provides practitioners with a direct role assignment for
the individual phases which gives structure to the process and thereby reduces the
remaining planning effort.
\end{quote}
\item[RQ2:] \emph{How is PQC applied when migrating software systems towards post-quantum cryptography?}
\begin{quote}
This research question aims to determine the actual adaptation of the software systems and standards for which the PQC\@ migration was carried out. Furthermore, we want to identify the usage of hybrid solutions after the migration.
\end{quote}
\item[RQ3:] \emph{Which challenges emerge when migrating software systems towards
post-quantum cryptography?}
\begin{quote}
Potential challenges play a decisive role for improving the planning and
implementation of the migration process. RQ3 therefore aims to
identify technical and organizational challenges.
\end{quote}
\end{description}
These research questions comprehensively address PQC migration. The RQs cover the
main building blocks and obstacles for a feasible and sensible migration, covering
the topic from different perspectives.

\subsection{Research Protocol Definition}
\label{s:methodology:protocol_definition}
The research protocol consists of our selection
criteria~\ref{s:methodology:selection_criteria}, our search
strategy~\ref{s:methodology:search_strategy}, and the quality assurance
plan~\ref{s:methodology:quality_assurance_plan}.

\subsubsection{Selection Criteria}
\label{s:methodology:selection_criteria}
Our selection criteria are divided into inclusion and exclusion criteria and
summarized in Table~\ref{tab:selection_criteria}. The acceptance of an article
in the final paper set required the fulfillment of all inclusion criteria
as well as the non-fulfillment of all exclusion criteria.

Inclusion criteria~1 and~2 as well as exclusion criteria 1~to~5 are generic criteria,
whereas all other criteria are specific to our research questions. The first
inclusion criterion aims to only include papers that provide new insights and,
therefore, exclude secondary studies such as surveys or similar. The second and
third inclusion criteria as well as exclusion criteria 1~to~5 are self-explanatory.
The term ``migration mechanisms'', mentioned in exclusion criteria~6 and~7,
generalizes various migration-related aspects like migration methodology, procedures,
frameworks, technologies, or similar. Exclusion criterion~8 focuses on migrating
existing standards towards post-quantum cryptography or designing a post-quantum
cryptography algorithm. Exclusion criterion~9 describes approaches for quantum key
distribution or similar topics.

\begin{table}[tb]
\small
    \caption{Selection criteria}
    \label{tab:selection_criteria}
    \raggedright
    \begin{tabularx}{\linewidth}{@{}X@{}}
      \toprule
      \textbf{Inclusion Criteria}\\
      \midrule
      \begin{enumerate}
        \item Primary studies or secondary studies that also provide an explicit
          research character.
        \item Studies that are classified as academic literature or specific grey
          literature, including technical and research reports, work in progress,
          white papers, government documents, conference \slash~workshop contributions,
          journal papers, PhD and master theses, as well as preprints.
        \item Studies that address methodologies, frameworks, or techniques for
          migrating existing software resources towards post-quantum cryptography.
      \end{enumerate} \\
      \toprule
      \textbf{Exclusion Criteria} \\
      \midrule
      \begin{enumerate}
        \item Duplicates of already included studies.
				\item Grey literature that are blog posts, websites, RSS feed, videos,
					podcasts, and webinars.
        \item Older version of an already included study.
				\item Studies that are not available, and hence not analyzable (e.g., the
					full text of a scientific article is not accessible or the link to a web
					page is broken).
        \item Studies written in any language other than English.
        \item Studies that do not address migration mechanisms.
				\item Studies that do not provide sufficient details about migration
					mechanisms.
        \item Studies that do not address a software system related migration.
        \item Studies that focus on quantum cryptography, i.e.,
          Quantum Key Distribution (QKD) or Quantum Communication.
      \end{enumerate} \\
      \bottomrule
    \end{tabularx}
  \end{table}

\subsubsection{Search Strategy}
\label{s:methodology:search_strategy}
The search strategy includes the selection of resources to be searched, the
definition of a search string, and the specification of a search procedure. Seven
different databases, one search engine, and 25 different conference proceedings were
selected as resources. These are shown in
Table~\ref{tab:resources_to_be_searched_table}.

\begin{lstlisting}[label=e:methodology:search_string, frame=single, caption=Search
	String, float, backgroundcolor=\color{lightgray!45}]
(migrat* OR shift* OR retrofit* OR
integrat* OR incorporat* OR mak*) AND
((quantum* AND crypto*) OR PQC)
\end{lstlisting}

The search string shown in Listing~\ref{e:methodology:search_string} was used for the databases and search enginge, refined iteratively, and validated according to our quality assurance plan. The search has
been further specified for each resource by using appropriate filters in combination with the search string to ensure accurate usage and therefore optimal
results. Our replication package contains the corresponding search string and its filter for each resource.

The main steps of our search procedure include the specification of an
initial upper limit for each resource, the removal of duplicates, the application of
the selection criteria, and the implementation of snowballing. The exact procedure is
described in Section~\ref{s:methodology:search_and_selection}.
\begin{table}
\small
  \caption{Resources we searched for relevant sources.}
  \label{tab:resources_to_be_searched_table}
  \raggedright
  \begin{tabularx}{\linewidth}{@{}X@{}}
    \toprule
    \textbf{Databases} \\
    \midrule
    \begin{enumerate}[nosep]
      \item ACM Digital Library
      \item IEEE Xplore
      \item Springer Link
      \item TNO Repository
      \item ETSI
      \item Cryptology ePrint Archive
      \item Open Quantum Safe Project
    \end{enumerate} \\
    \toprule
    \textbf{Search Engines} \\
    \midrule
    \begin{enumerate}[nosep]
      \item Google Scholar
    \end{enumerate} \\
    \toprule
    \textbf{Conference Proceedings} \\
    \midrule
    \begin{enumerate}[nosep]
      \item Advances in Cryptology
      \item International Conference on the Theory and Application of Cryptographic
        Techniques
      \item IEEE Symposium on Security and Privacy
      \item IEEE European Symposium on Security and Privacy
      \item Usenix Network and Distributed System Security Symposium
      \item ACM Conference on Computer and Communications Security
      \item European Symposium On Research In Computer Security
      \item International Conference on Applied Cryptography and Network Security
      \item International Conference on Practice and Theory in Public Key
        Cryptography
      \item International Conference on Information Security and Cryptography
      \item Selected Areas in Cryptography
      \item IMA International Conference on Cryptography and Coding
      \item IEEE Symposium on Foundations of Computer Science
      \item International Conference on Availability, Reliability and Security
      \item Post-quantum cryptography
      \item International Symposium on Foundations \& Practice of Security
      \item IEEE International Conference on Pervasive Computing and Communications
      \item IEEE Computer Security Foundations Symposium
      \item IEEE International Symposium on Network Computing and Applications
      \item IEEE Consumer Communications and Networking Conference
      \item IEEE Symposium on Computers and Communications
      \item ACS/IEEE International Conference on Computer Systems and Applications
      \item ACM International Conference on Emerging Networking Experiments and
        Technologies
      \item ACM Symposium on Applied Computing
      \item International Conference on the Theory and Application of Cryptology and
        Information Security
    \end{enumerate} \\
    \bottomrule
  \end{tabularx}
\end{table}

\subsubsection{Quality Assurance Plan}
\label{s:methodology:quality_assurance_plan}
The last element of our protocol is the quality assurance plan. It has the goal to
prevent potential bias by ensuring various researchers engage in the whole
process and that there is a common understanding of the underlying principles.
Table~\ref{tab:quality_assurance_plan_table} shows the different
assurances and the steps taken.
\begin{table}
\small
  \caption{Quality assurance plan.}
  \label{tab:quality_assurance_plan_table}
  \raggedright
  \begin{tabularx}{\linewidth}{@{}X@{}}
    \toprule
    \textbf{Definition of Research Questions} \\
    \midrule
    \begin{enumerate}[nosep]
      \item The first author defines the research questions
      \item All other authors review the research questions
      \item They discuss disagreements until they reach a consensus
    \end{enumerate} \\
    \toprule
    \textbf{Definition of Search Strategy Procedure} \\
    \midrule
    \begin{enumerate}[nosep]
			\item The first author defines the search strategy including the search engine,
				databases, conference proceedings, as well as the search terms and steps
      \item The other authors review the search strategy
      \item They discuss disagreements until they reach a consensus
    \end{enumerate} \\
    \toprule
    \textbf{Definition of Inclusion and Exclusion Criteria} \\
    \midrule
    \begin{enumerate}[nosep]
			\item The first author selects inclusion and exclusion criteria based on the
				reference set and research questions
      \item The second and third author review the criteria
      \item All authors adjust the criteria until they reach a consensus
    \end{enumerate} \\
    \toprule
    \textbf{Definition of Data Extraction Criteria} \\
    \midrule
    \begin{enumerate}[nosep]
      \item The first author selects the data extraction criteria
      \item The second and third author review the criteria
      \item All authors adjust the criteria until they reach a consensus
    \end{enumerate} \\
    \toprule
    \textbf{Selection of Primary Studies} \\
    \midrule
    \begin{enumerate}[nosep]
      \item The first author selects primary studies
      \item The second author independently verifies that the defined search strings
				yielded all papers of the reference set
	  \end{enumerate} \\
    \toprule
    \textbf{Random Assessment of Included \slash~Excluded Publications} \\
    \midrule
    \begin{enumerate}[nosep]
      \item Select a random sample (roughly 10\%, depends on the overall amount of
        papers) of the search results
      \item The second author applies the inclusion and exclusion criteria
        independently on the selected publications
      \item Compare the results to the results of the first author
      \item They discuss disagreements until they reach a consensus
      \item If both authors consent to the selection in most cases they proceed with
				the process, otherwise they repeat this step
		\end{enumerate}
      If there is still disagreement discuss a consensus. Then, the first
      author applies the in- and exclusion criteria on the whole search result.
      Afterwards, start with point 2 again.\\~\\
    \toprule
    \textbf{Development of a Common Data Extraction Understanding} \\
    \midrule
    \begin{enumerate}[nosep]
      \item The first two authors choose a paper of the reference set
      \item Both authors perform the data extraction
      \item Both authors compare their results of the data extraction
      \item They discuss disagreements until they reach a consensus
      \item Redo step 1-4 for two more papers of the reference set
    \end{enumerate} \\
		\midrule
  \end{tabularx}
\end{table}
\begin{table}
\small
  \begin{tabularx}{\linewidth}{@{}X@{}}
    \toprule
    \textbf{Random Quality Check of Data Extraction} \\
    \midrule
    \begin{enumerate}[nosep]
      \item Select a random sample (roughly 10\%, depends on the overall amount of
        papers) of included publications
      \item The second author checks the quality of the data extraction in accordance
        with the research questions
      \item Compare the results to the result of the first author
      \item Either both authors agree on the extraction in general or they find a new
				consensus and then the first author manually examines the rest of the publications (90\%) afterwards
    \end{enumerate} \\
    \bottomrule
  \end{tabularx}
\end{table}

\subsection{Reference Set}
\label{s:methodology:reference_set}
The reference set serves as a guideline to identify articles that could be part
of the final paper set. It is also used to formulate the inclusion and exclusion
criteria which are then tested against the reference set. Accordingly, the reference
set also changed until the criteria were final. The initial reference set is based on:
\begin{enumerate}
    \item the referenced articles in related secondary studies;
	\item our own pre-existing knowledge as well as knowledge from existing networks of
			experts in the PQC field;
    \item searches in Google Scholar.
\end{enumerate}
The reference set was used to validate the outcomes of the search and selection
step.

\subsection{Search and Selection}
\label{s:methodology:search_and_selection}
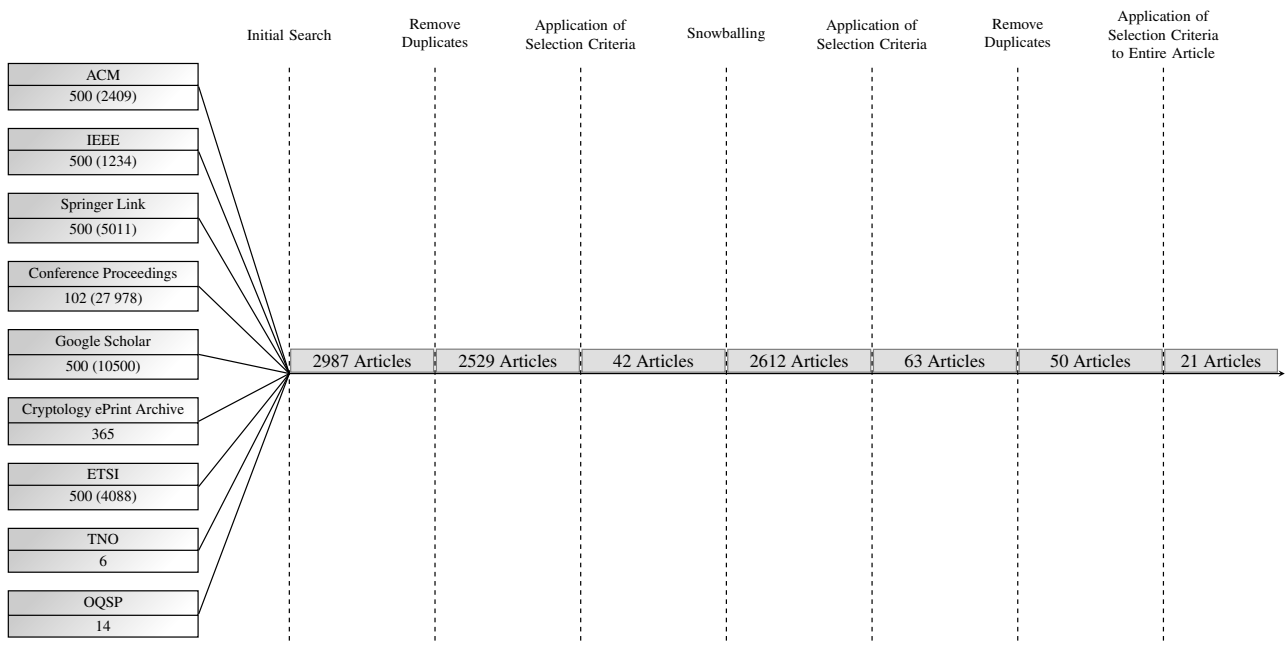
\begin{figure*}
	\centering
	\begin{tikzpicture} [
  sta/.style={state, align=center, drop shadow},
  angle/.style={top color=gray!40, shading angle=45},
	box/.style={draw=black, anchor=west, angle, align=center, minimum width=4.38cm},
  component/.style={rectangle, draw=gray, fill=lightgray!50, minimum width=3.3cm,
	yshift=0.018cm},
	partbox/.style={box, rectangle split, rectangle split parts=2},
	lane/.style={text width=2.7cm, align=center, minimum height=1.5cm},
  blue_box/.style={top color=blue!35, bottom color=blue, box},
  red_box/.style={top color=red!35, bottom color=red, box},
  green_box/.style={top color=green!35, bottom color=green, box},
  gray_box/.style={top color=gray!35, bottom color=gray, box},
  arrow/.style={draw=black, line width=3, stealth},
	thick,
	>=stealth,
	auto,
	every initial by arrow/.style={*->},
	initial distance=0.6cm,
	initial text={},
	node distance=0.4cm
]

\node[xshift=6.5cm, lane] (init) {Initial Search};
\node[right=of init, lane] (dupes) {Remove\\Duplicates};
\node[right=of dupes, lane] (sel) {Application of Selection Criteria};
\node[right=of sel, lane] (snow) {Snowballing};
\node[right=of snow, lane] (sel2) {Application of Selection Criteria};
\node[right=of sel2, lane] (dupes2) {Remove\\Duplicates};
\node[right=of dupes2, lane] (sel_entire) {Application of Selection Criteria to
	Entire Article};

\node[partbox, yshift=-1.2cm] (acm) {ACM\nodepart{two} 500 (2409)};
\node[partbox, below=of acm] (ieee) {IEEE\nodepart{two} 500 (1234)};
\node[partbox, below=of ieee] (spring) {Springer Link\nodepart{two} 500 (5011)};
\node[partbox, below=of spring] (conf) {Conference Proceedings\nodepart{two} 102 (27
978)};
\node[partbox, below=of conf] (g_schol) {Google Scholar\nodepart{two} 500 (10500)};
\node[partbox, below=of g_schol] (eprint) {Cryptology ePrint Archive\nodepart{two} 365};
\node[partbox, below=of eprint] (etsi) {ETSI\nodepart{two} 500 (4088)};
\node[partbox, below=of etsi] (tno) {TNO\nodepart{two} 6};
\node[partbox, below=of tno] (oqsp) {OQSP\nodepart{two} 14};

\coordinate (mid) at ([yshift=-7.8cm] init);
\draw (acm.east) -- (mid);
\draw (ieee.east) -- (mid);
\draw (spring.east) -- (mid);
\draw (conf.east) -- (mid);
\draw (g_schol.east) -- (mid);
\draw (eprint.east) -- (mid);
\draw (etsi.east) -- (mid);
\draw (tno.east) -- (mid);
\draw (oqsp.east) -- (mid);

\draw[->] (mid) -- ++(3.3625cm, 0) node[pos=0.5, align=center, component] {\large2987
    Articles}
	-- ++(3.3625cm, 0) node[pos=0.5, align=center, component] {\large2529 Articles}
	-- ++(3.3625cm, 0) node[pos=0.5, align=center, component] {\large42 Articles}
	-- ++(3.3625cm, 0) node[pos=0.5, align=center, component] {\large2612 Articles}
	-- ++(3.3625cm, 0) node[pos=0.5, align=center, component] {\large63 Articles}
	-- ++(3.3625cm, 0) node[pos=0.5, align=center, component] {\large50 Articles}
	-> ++(2.8cm, 0) node[pos=0.475, align=center, component, minimum width=2.6cm]
	{\large21 Articles};
\draw[dashed] (init) -- ++(0,-14.1cm);
\draw[dashed] (dupes) -- ++(0,-14.1cm);
\draw[dashed] (sel) -- ++(0,-14.1cm);
\draw[dashed] (snow) -- ++(0,-14.1cm);
\draw[dashed] (sel2) -- ++(0,-14.1cm);
\draw[dashed] (dupes2) -- ++(0,-14.1cm);
\draw[dashed] (sel_entire) -- ++(0,-14.1cm);

\node [xshift=2.25cm, right=of sel_entire] {};
\end{tikzpicture}
	\caption{Conducting the Search and Selection}
	\label{fig:search_conduction}
\end{figure*}
In the course of the search and selection process, a total of 51,605 hits were
retrieved from the resources specified in
Table~\ref{tab:resources_to_be_searched_table} by applying our search
strings (see Listing~\ref{e:methodology:search_string}). We set the number of items from
each resource to 500, as it turned out that there were no other relevant matches
regarding the specified search term found after this limit.

We made an exception to this approach for conference proceedings, where we conducted a manual
search without a predefined search string on the corresponding conference pages of \emph{DBLP}~\footnote{https://dblp.org/}. We selected the conferences based on the \emph{CORE ranking
portal}~\footnote{https://www.core.edu.au/},
which provides a quality assessment for conferences.

We chose 25 conferences (see Table~\ref{tab:resources_to_be_searched_table}), ranging from A* to C ratings.
Out of a total of 27,978 conference findings, we deemed 102 relevant based on our
search string. The analysis of conference proceedings was restricted to the last 12
years, ensuring the inclusion of all articles published from 2012 onwards.

Figure~\ref{fig:search_conduction} illustrates the search and selection
procedure, which is explicated below. Following the initial search that yielded 2,987
articles, duplicates were removed. Subsequently, the inclusion and exclusion criteria
were first applied to the titles and keywords of the remaining 2,593 articles. Then,
selection criteria were applied to the abstract and conclusion followed by applying
them to the overall article structure, i.e., to their section titles, in the
subsequent step. Backward snowballing was conducted next, incorporating all
references of the remaining articles, with the goal of discovering additional
relevant articles beyond our initially searched resources. The selection criteria
were again applied to the articles added through snowballing, followed by the removal
of duplicates. As a final step, the selection criteria were applied to the entire
text, necessitating a comprehensive reading of each individual article. This
comprehensive process resulted in the creation of the final paper set, consisting of 21 papers.

\subsection{Data Extraction and Synthesis}
\label{s:methodology:data}
Once we had identified the relevant publications, we extracted and analysed
the data contained in them systematically using thematic coding.
The \emph{data extraction categories} we defined are described in
Section~\ref{s:methodology:data_extraction_categories} and used for the actual
\emph{data extraction} detailed in Section~\ref{s:methodology:data_extraction}.
Finally, we describe the \emph{data analysis} and \emph{synthesis} in
Section~\ref{s:methodology:data_synthesis}.

\subsubsection{Data Extraction Categories}
\label{s:methodology:data_extraction_categories}
Data extraction categories allow us to map content to the research questions.
We used a mixture of predefined categories and emergent categories
identified during thematic coding. In the following, we present the final
set of categories used to code the papers in the final set.

Multiple categories are assigned to each RQ and are recorded in Tables~\ref{t:rq_1},~\ref{t:rq_2},
and~\ref{t:rq_3}. Table~\ref{t:rq_1} shows the categories Diagnosis, Plan, Execution,
and Maintenance for RQ1 which, from the literature's and our point of view, reflect
the four central phases of migration. Table~\ref{t:rq_2} shows the technical
applications of PQC that have played a key role in the migration so far. It shows which
software systems have been migrated, which standards should be and have been adapted,
as well as the extent to which hybrid migration plays a role in the literature. Last
but not least, Table~\ref{t:rq_3} lists the challenges that arise
during such a migration.
\begin{table}
\small
  \renewcommand{\cellalign}{tl}
  \centering
  \caption{Data extraction criteria corresponding to research question RQ1.
    \label{t:rq_1}}
  \begin{tabular}{ll}
    \toprule
    \multicolumn{2}{c}{\textbf{Migration Steps (RQ1)}} \\
    \midrule
    Diagnosis & Asset Identification \\
    & Risk Assessment \\
    & Cryptographic Identification \\
    & Cryptographic Assessment \\
    \midrule
    Plan & Cryptographic Prioritization \\
    & Migration Plan \\
    \midrule
    & Execution \\
    \midrule
    & Maintenance \\
    \bottomrule
  \end{tabular}
\end{table}

\begin{table}
 \small
 \centering
  \caption{Data extraction criteria corresponding to research question RQ2.
    \label{t:rq_2}}
  \begin{tabular}{ll}
    \toprule
    \multicolumn{2}{c}{\textbf{Applications of PQC (RQ2)}} \\
    \midrule
    Hybrid Migration & Claimed \\
    & Recommended \\
    & Mentioned \\
    \midrule
    Migrated Software System & Nginx \\
    & DB2 \\
    & PostgreSQL \\
    & IoT Infrastructure \\
    & Enterprises \\
    & ROS \\
    & Blockchain \\
    & Messenger \\
    & CA \\
    & Microservices \\
    \midrule
    Adjusted Technology & IBE \\
    & TLS \\
    & SSH \\
    & S/MIME \\
    & PGP \\
    & X.509 \\
    & OCSP \\
    & CMP \\
    & EST \\
    & IPsec \\
    \bottomrule
  \end{tabular}
\end{table}

\begin{table}
\small
  \renewcommand{\cellalign}{tl}
  \centering
  \caption{Data extraction criteria corresponding to research question RQ3.
    \label{t:rq_3}}
  \begin{tabular}{ll}
    \toprule
    \multicolumn{2}{c}{\textbf{Challenges (RQ3)}} \\
    \midrule
    Organizational & Education \\
	& Skeptics \\
	& Resource Planning \\
    \midrule
    \multirow{2}{*}{\makecell{Missing \slash~Faulty PQC \\ Features, Guidelines or
	  Support}} & Performance Issues \\
	& System Downtime \\
	& HW Resources \\
    \midrule
    Code Base \& Documentation & Large \& Complex \\
	& Hard Coding \\
	& Legacy Encryption \\
	& Legacy Environment \\
	& In Flux \\
	& Large Configuration \\
    \midrule
    Cryptographic Inventory & \\
    \bottomrule
  \end{tabular}
\end{table}

\subsubsection{Data Extraction}
\label{s:methodology:data_extraction}
The objective of this activity is to extract all relevant information to answer
our research questions. The loop shown in the process described in 
Section~\ref{s:methodology:process} symbolizes 
that we conducted this process iteratively. We did the final thematic coding
of all papers with the final set of categories by marking and annotating the
PDFs of the papers in the final paper set.
We then extracted all coded information into an online whiteboard.

\subsubsection{Data Analysis and Synthesis}
\label{s:methodology:data_synthesis}
We performed a vertical data analysis with the intention of assigning the extracted
data to the corresponding extraction categories. For this purpose, we carried out
a content analysis by collecting, examining, and summarizing the data and trying to
identify trends. The quantitative part of the analysis involved providing descriptive
statistics to ensure a better understanding. For the qualitative analysis, we analyzed 
the information by grouping extracts that belong together. Within these groups, we
then identified similarities and differences. We used this information to identify
possible patterns and trends. These findings are 
presented in detail in Section~\ref{s:results}.

\section{Results}
\label{s:results}
In this section, we present an analysis of the selected articles. First, we provide
an overview of descriptive statistics of the articles in
Section~\ref{sec:results:descriptivestatistics}. Subsequently, we will answer our
three research questions in Section~\ref{sec:results:rq1} (migration steps),
Section~\ref{sec:results:rq2} (technological modifications), and
Section~\ref{sec:results:rq3} (challenges) respectively.

\subsection{Descriptive Statistics}
\label{sec:results:descriptivestatistics}
\begin{table*}
\small
  \renewcommand{\cellalign}{tl}
  \centering
  \caption{Classification of all papers according to our descriptive statistics
    criteria.\label{t:descriptive_stats}}
  \begin{tabular}{c!{\color{lightgray}\vrule}ccc!{\color{lightgray}\vrule}ccc!{\color{lightgray}\vrule}ccc!{\color{lightgray}\vrule}cc!{\color{lightgray}
	\vrule}cc!{\color{lightgray}\vrule}cc}
    \toprule
    \textbf{Ref} & \multicolumn{3}{c!{\color{lightgray}\vrule}}{\textbf{Literature}} &
	  \multicolumn{3}{c!{\color{lightgray}\vrule}}{\textbf{Author Affiliation}} &
	  \multicolumn{3}{c!{\color{lightgray}\vrule}}{\textbf{Validator}} &
	  \multicolumn{2}{c!{\color{lightgray}\vrule}}{\textbf{Contribution}} &
	  \multicolumn{2}{c!{\color{lightgray}\vrule}}{\textbf{Realization}} &
	  \multicolumn{2}{c}{\textbf{Focus}}\\
    \midrule
    & \rotatebox[origin=l]{90}{Peer-Reviewed} &
      \rotatebox[origin=l]{90}{Preprint} &
      \rotatebox[origin=l]{90}{Gray} &
      \rotatebox[origin=l]{90}{Academia} &
      \rotatebox[origin=l]{90}{Industry} &
      \rotatebox[origin=l]{90}{\parbox{2cm}{Stan\-dar\-di\-zation Body}} &
      \rotatebox[origin=l]{90}{3rd Party} &
      \rotatebox[origin=l]{90}{Committee} &
      \rotatebox[origin=l]{90}{Author(s)} &
      \rotatebox[origin=l]{90}{Guideline} &
      \rotatebox[origin=l]{90}{Framework} &
      \rotatebox[origin=l]{90}{Illustration} &
      \rotatebox[origin=l]{90}{Exemplar} &
      \rotatebox[origin=l]{90}{Organizational} &
      \rotatebox[origin=l]{90}{Technical} \\
	\midrule
    \citer{caraf} & \textbullet &  &  &  & \textbullet &  & \textbullet &
	  &  &  & \textbullet & \textbullet &  & \textbullet &  \\
    \citer{cyber} &  &  & \textbullet &  &  & \textbullet &  & \textbullet &
	   & \textbullet &  & \textbullet &  & \textbullet &  \\
    \citer{impact_hybrid} & \textbullet &  &  & \textbullet & \textbullet &
	  &  \textbullet&  &  & \textbullet &  &  & \textbullet &  & \textbullet \\
    \citer{microservice} &  &  & \textbullet & \textbullet &  &  &  &  &
	  \textbullet & \textbullet &  &  & \textbullet &  & \textbullet \\
    \citer{tls_database} & \textbullet &  &  & \textbullet & \textbullet &
	  & \textbullet &  &  & \textbullet &  &  & \textbullet &  & \textbullet \\
    \citer{ibe} &  & \textbullet &  & \textbullet &  &  &  &  & \textbullet
	  & \textbullet &  &  & \textbullet &  & \textbullet \\
    \citer{ibm_db2} & \textbullet &  &  & \textbullet & \textbullet &  &
	  \textbullet &  &  & \textbullet &  &  & \textbullet & \textbullet &
	  \textbullet \\
    \citer{dep_analysis} &  & \textbullet &  & \textbullet & \textbullet &
	  &  &  & \textbullet &  & \textbullet & \textbullet & \textbullet & \textbullet
	  & \textbullet \\
    \citer{nist_migration} &  &  & \textbullet &  &  & \textbullet &  &
	  \textbullet &  & \textbullet &  & \textbullet &  & \textbullet &  \\
    \citer{making_decisions} &  &  & \textbullet &  &  & \textbullet &  &
	  \textbullet &  & \textbullet &  & \textbullet &  & \textbullet &  \\
    \citer{nginx_migration} &  &  & \textbullet & \textbullet &  &  &  &
	  & \textbullet & \textbullet &  &  & \textbullet &  & \textbullet \\
    \citer{hbs_iot} & \textbullet &  &  & \textbullet &  &  & \textbullet &
	  &  & \textbullet &  & \textbullet &  &  & \textbullet \\
    \citer{pmmp} &  & \textbullet &  & \textbullet &  &  &  &  &
	  \textbullet &  & \textbullet & \textbullet &  & \textbullet &  \\
    \citer{pqc_ros} & \textbullet &  &  &  & \textbullet &  & \textbullet &
	  &  &  & \textbullet &  & \textbullet &  & \textbullet \\
    \citer{pqfabric} & \textbullet &  &  & \textbullet & \textbullet &  &
	  \textbullet &  &  &  & \textbullet &  & \textbullet &  & \textbullet \\
    \citer{pqc_email} & \textbullet &  &  & \textbullet & \textbullet &  &
	  \textbullet &  &  &  & \textbullet &  & \textbullet &  & \textbullet \\
    \citer{cits} &  &  & \textbullet &  &  & \textbullet &  & \textbullet &
	  & \textbullet &  & \textbullet &  & \textbullet &  \\
    \citer{refine_mosca} &  &  & \textbullet &  &  & \textbullet &  &
	  \textbullet &  &  & \textbullet & \textbullet &  & \textbullet &  \\
    \citer{complex_path} &  &  & \textbullet & \textbullet &  &  &  &  &
	  \textbullet & \textbullet &  & \textbullet &  & \textbullet &  \\
    \citer{handbook} &  &  & \textbullet & \textbullet & \textbullet &
	  \textbullet &  & \textbullet &  &  & \textbullet & \textbullet &  &
	  \textbullet &  \\
    \citer{ibm_z} &  &  & \textbullet &  & \textbullet &  &  &  &
	  \textbullet & \textbullet &  &  & \textbullet & \textbullet & \textbullet \\
		\midrule
		\midrule
		$\sum$ & 8 & 3 & 10 & 13 & 10 & 6 & 8 & 6 & 7 & 13 & 8 & 11 & 11 & 12 & 12 \\
	\bottomrule
  \end{tabular}
\end{table*}

In this section we present the descriptive statistics of the articles with the help
of Table~\ref{t:descriptive_stats}.

\subsubsection{Literature}
\label{sec:results:literature}
The first criterion is the type of literature of the 21 selected articles. This gives
us an initial impression of the scientific quality of the
articles. We differentiate between preprints, peer-reviewed, and gray literature.

All articles that are neither peer-reviewed nor preprints belong to
gray literature. The main reason for the introduction of preprints as a subcategory
lies in the special nature of manuscripts intended for a scientific audience. In
contrast to other forms of gray literature, a preprint is characterized by a specific
writing style tailored to scientific communication. Ten of the 21 articles are gray
literature, three articles are currently available as preprints, and the remaining
eight articles have been peer-reviewed.

\subsubsection{Author Affiliation}
\label{sec:results:author_affiliation}
The credibility and potential bias of the extracted data are significantly influenced
by the authors' affiliations. We established the three
categories academia, industry, and standardization body (e.g., European
Telecommunications Standards Institute) as suitable criteria. Within the literature,
instances exist, as exemplified by~\citer{handbook}, where authors span all three
categories. Conversely, articles such as~\citer{nist_migration}
or~\citer{nginx_migration} illustrate cases where authors exclusively affiliate with
standardization bodies or universities. In 13 of the present articles, some of the
authors are employed in academia, while in ten articles several authors belong
to the industry. In six articles, some of the authors belong to standardization
organizations.

\subsubsection{Validator}
\label{sec:results:validator}
We define the term validator as the entity incorporated into a paper's methodology to
authenticate and validate its results. We differentiate between 3rd party,
committee, and the authors of the article. By 3rd party we mean a group of experts
and researchers in the field who are independent of the authors. This criterion is
met for all eight peer-reviewed articles. Committee refers to a group of experts in
the field who, in contrast to 3rd party, do not have to be independent. This
means that the authors and other contributors can also be part of the committee. The
final instance are the authors themselves, who validate their own results.
Six articles are assigned to the committee and seven to the authors.

\subsubsection{Contribution}
The contribution criterion categorizes the various articles according to their scope
as either guidelines or frameworks. In our case, frameworks describe the content in
detail, whereas guidelines present the content briefly and concisely. An example of a
framework is Attema et al.~\citer{handbook}, who describe the process, personas, and
technology of the migration in great detail. ETSI's article~\citer{cyber}, on the
other hand, presents the migration steps, roles, and technical contexts in a brief
and concise manner and was therefore classified as a guideline.

\subsubsection{Realization}
According to our definition, an illustration serves as a
conceptual representation of migration, while an exemplar offers insights
predominantly from the viewpoint of practical execution.

Tueno et al.~\citer{tls_database} and Zhang et al.~\citer{ibm_db2} are examples for
papers categorized as exemplars, as the focus of both articles is on the concrete
execution of the migration by using a selected software system as a use case. Ma et
al.~\citer{caraf} and von Nethen et al.~\citer{pmmp} illustrate the migration and the
associated aspects but do not apply them to a specific case. In our opinion,
Hasan et al.~\citer{dep_analysis} is the only article that focuses on both the
illustrative and the exemplary part in sufficient detail.

\subsubsection{Focus}
Using ``focus'', we categorize articles by whether they describe the migration from an
organizational perspective or from a technical perspective. With organizational
focus, the process, roles, as well as challenges, costs, and effort of the migration
are considered in particular, as is the case with von Nethen et al.~\citer{pmmp}.
With technical focus, the technical standards (e.g., protocols), the software
systems, and the technical modifications are explained in detail and might be
implemented in a concrete use case. Apart from Zhang et al.~\citer{ibm_db2} and
Hasan et al.~\citer{dep_analysis}, the remaining articles primarily describe the
migration from either an organizational or technical perspective.

\subsection{RQ1: Migration Steps}
\label{sec:results:rq1}
We find that PQC migration is structured into several steps. Additionally, most reviewed papers define a \emph{migration team}
consisting of experts and managers of the to be migrated software system or infrastructure. This team
drives the migration process and cares for its completion.

\subsubsection{Migration Step Terminology}
\label{sec:results:rq1:stepsterminology}
The literature neither provides a common understanding of how fine-grained the
definition of the migration steps should be, nor about their terminology. This makes
a comparison of their respective migration processes difficult. Thus, we first define our
migration step terminology. Afterwards in Section~\ref{sec:results:rq1:literaturesteps}, we map the
literature to our definitions, thus validating our schema. We concluded that
the following definitions fit the overall literature best:
\begin{description}
	\item[Diagnosis] is the first phase where the migration team collects
		information about the infrastructure. It is necessary to plan and conduct the
		migration.
	\item[Planning] comprises the planning of the migration process considering
		the previously collected information.
	\item[Execution] consists of the actual migration, ideally according to the
		previously generated plan.
	\item[Maintenance] is an ongoing phase starting together with the
		execution. It is motivated by the fact that the considered infrastructure will
		change over time, both during the execution phase and afterwards. Thus,
	    the inventories must be adapted over time and the infrastructure adjusted.
\end{description}
For a more detailed analysis, we divide the phases into separate concrete steps
according to what is reported in the surveyed literature.

\emph{Diagnosis} comprises the following:
\begin{description}
	\item[Asset Identification] is the step in which all assets of an
		infrastructure are identified. The result is a structured artefact (e.g. a list) of all relevant
		assets of the enterprise, also called an \emph{asset inventory}.
	\item[Risk Assessment] comprises an analysis of all assets regarding the
    cost or damage a potential attacker could cause and the chances of their success.
    This also feeds into the asset inventory or can be a separate artefact.
	\item[Cryptographic Identification] is the step where all the assets'
		cryptography, e.g., which hash functions, symmetric and asymmetric encryption, or
		certificates and their implementations, is collected and recorded in a
		\emph{cryptographic inventory}.
	\item[Cryptographic Assessment] is the last preparation step and consists
		of an analysis of the applied cryptography, especially in the context of a
		quantum threat. The result enhances the cryptographic inventory.
\end{description}
With the asset and cryptographic inventory as well as the risk analysis at hand, we
can now start the \emph{Planning Phase} with the following steps:
\begin{description}
	\item[Cryptographic Prioritization] gives each asset and its cryptography a
		priority, considering its corresponding risk assessment result and potentially
		also other factors such as cost of migration or external dependencies. This
		information is compiled into the existing inventories.
	\item[Migration Plan] is the step in which the team creates the actual plan
		and timeline for the migration. The result is a document containing at least the
		information who migrates what and when.
\end{description}
We expect the \emph{Execution and Maintenance Phases} to be complex and tedious. Yet,
their dynamic nature and dependence on the previous phases' results makes it hard to
give them more static structure. The lack of detailed descriptions of these phases in
the literature is also an indicator that the research community can still benefit
from further research into this direction.

\subsubsection{Migration Steps in the Literature}
\label{sec:results:rq1:literaturesteps}
With the migration phases defined, we can now structure the literature according to
the steps it considered. This means we mapped the migration steps to our definitions
and terminology. Table~\ref{t:migration_steps} shows the results of the mapping and
which papers focussed on what topics. In the following paragraphs, we will describe
the differences in the definitions of the steps' in the literature compared to ours.
\begin{table*}
\small
  \centering
  \caption{The migration steps according to our definitions and how the literature
    covers them. We removed all papers which did not address the migration procedure.
	  \label{t:migration_steps}}
  \begin{tabular}{c!{\color{lightgray}\vrule}cccc!{\color{lightgray}\vrule}
		cc!{\color{lightgray}\vrule}c!{\color{lightgray}\vrule}c}
  \toprule
    \textbf{Ref} & \multicolumn{4}{c!{\color{lightgray}\vrule}}{\textbf{Diagnosis}} &
      \multicolumn{2}{c!{\color{lightgray}\vrule}}{\textbf{Plan}} &
      \textbf{Execution} & \textbf{Maintenance} \\ \midrule
    & \rotatebox[origin=l]{60}{Asset identification} &
      \rotatebox[origin=l]{60}{Risk assessment} & \rotatebox[origin=l]{60}
      {\parbox{2cm}{Cryptographic\\ identification}} & \rotatebox[origin=l]{60}{\parbox{2cm}{Cryptographic\newline
      assessment}} & \rotatebox[origin=l]{60}{\parbox{2cm}{Cryptographic Prioritization}} &
      \rotatebox[origin=l]{60}{Migration Plan} &  &  \\
    \midrule
	\citer{caraf} & \textbullet & \textbullet & \textbullet & \textbullet &
      & \textbullet & & \\
	\citer{cyber} & \textbullet & \textbullet & \textbullet & \textbullet &
      \textbullet & \textbullet & \textbullet &  \\
	\citer{impact_hybrid} &  &  &  &  &  &  & \textbullet &  \\
	\citer{ibm_db2} & \textbullet &  &  &  & \textbullet & \textbullet &
      \textbullet & \textbullet \\
	\citer{dep_analysis} & \textbullet & \textbullet & \textbullet &
      \textbullet &  &  & \textbullet & \textbullet \\
	\citer{nist_migration} & \textbullet & \textbullet & \textbullet &
      \textbullet & \textbullet & \textbullet &  &  \\
	\citer{making_decisions} & \textbullet &  & \textbullet & \textbullet &
      & & &\\
	\citer{hbs_iot} &  &  &  &  &  & \textbullet &  &  \\
	\citer{pmmp} & \textbullet &  & \textbullet &  &  & \textbullet &
      \textbullet & \textbullet\\
	\citer{pqc_email} &  &  &  &  &  &  & \textbullet &  \\
	\citer{refine_mosca} &  &  &  &  & \textbullet &  &  &  \\
	\citer{complex_path} &  & \textbullet & \textbullet &  &  & \textbullet
      &  & \textbullet\\
	\citer{handbook} & \textbullet & \textbullet & \textbullet &
      \textbullet & \textbullet & \textbullet & \textbullet & \textbullet \\
	\citer{ibm_z} &  \textbullet & \textbullet & \textbullet & \textbullet &
      \textbullet & \textbullet & \textbullet &  \\
	\midrule
	\midrule
  $\sum$ & 9 & 7 & 9 & 7 & 6 & 9 & 8 & 5 \\
  \bottomrule
  \end{tabular}
\end{table*}

In general, the most documented migration step is \emph{Diagnosis}. Within
Diagnosis, most papers mention the compilation of the cryptographic inventory, but
provide very few details on how to do this practically. In addition, their definition
of what the cryptographic inventory should comprise
differs. Some papers see it as an inventory purely for cryptographic
assets~\citer{complex_path}. Other papers split the inventory into several
sub-inventories~\citer{dep_analysis, nist_migration, making_decisions, handbook,
ibm_z}, e.g., a data inventory. The ETSI~\citer{cyber} additionally differentiates
between an organization's data, its infrastructure, and its supplier
inventory~\citer{cyber}. Since the cryptographic inventory is a key artifact in any
PQC migration, this lack of concrete guidance is a serious research gap.

The reviewed literature showed more agreement regarding the \emph{Planning
Phase}. While the papers differ in terms of
granularity, the ETSI~\citer{cyber} stands out again. In addition to the
common prioritization and planning steps, it also gives advice on planning for
failure or disorderly transition. It mentions that some assets may need to be
isolated or at least kept in different zones, as migrating them (early) might not be
feasible. Finally, it discusses trust and key management during
migration~\citer{cyber}.

For the \emph{Execution Phase} the literature is rather sparse, possibly since there
is still a lack of experience regarding actual PQC migrations. The papers that
do cover the execution phase mostly stress the importance of transitional solutions
for successfully conducting the migration without increasing the overall risk. These
include the addition of pre-shared keys~\citer{handbook}, testing the compatibility
with just a small set of systems first~\citer{pmmp}, isolating hard-to-migrate assets
from the network~\citer{cyber}, and, most predominantly, using hybrid cryptography
schemes~\citer{impact_hybrid, handbook, ibm_z}.

The last phase is the \emph{Maintenance} or \emph{Remediation Phase}. It is also not
covered extensively in the literature, probably for the same reason as the
execution phase. The papers that do address this phase mention three aspects: 1) The
most prominent one is that an organization must not make the mistake to only allocate
resources for the migration itself without considering the maintenance of the new or
changed infrastructure~\citer{ibm_db2, dep_analysis, handbook}. 2) For management it
is usually important to have information about the success of the migration. Thus, it
is necessary to track the costs and measure the impact of the conducted migration
steps~\citer{ibm_db2, pmmp}. 3) Documenting the lessons learned in order to improve
the subsequent PQC migration steps and future migrations, especially since PQC is
currently a fast moving technology~\citer{ibm_db2}.

Seven papers do not address any of the migration phases of
Table~\ref{t:migration_steps}. Those papers focus on technical aspects of the
migration of specific software. Since they all migrated a specific system which is
not in productive use, the migration's organizational impact is no topic in
those papers. Instead, the authors usually chose a generic software development
approach. This includes analyzing the software, searching a way to solve the
given problem, hence, integrating PQC, implementing their preferred solution, and
finally evaluating their result, ideally against predefined metrics. Thus, these
papers do not give insight into the topic of how to manage the migration of complex
distributed systems or infrastructures which are currently in use. Still, we deem
them an important contribution to the literature about PQC migration, due to their
information about the challenges and possible solutions specific for developing
PQC-ready software.

\subsubsection{Roles in Migration}
\label{sec:results:rq1:roles}
The team driving the migration has to include several roles to have both the
expertise and the authority within the organization to successfully complete the
migration process. Most papers that do consider these roles define them including
their names, necessary skills, and competences. Only Muller and
Heesch~\citer{making_decisions} leave it at declaring their skills. We clustered the
literature's role definitions into four archetypes: 1) \emph{Migration Manager},
2) \emph{Asset \& Risk Manager}, 3) \emph{Security Expert}, 4) \emph{Developer or Administrator}.

Most of the literature that covers the roles involved in the migration process
mention at least one aspect of each role that is important for the migration team.
Table~\ref{t:role_coverage} shows the exact coverage of each paper regarding the role
archetypes.
\begin{table*}
\small
  \centering
  \caption{The role coverage of the reviewed papers regarding our archetypes. All
		papers that do not mention roles are not listed.\label{t:role_coverage}}
  \begin{tabular}{ccccc}
    \toprule
  \textbf{Ref} & \textbf{Migration Manager} & \textbf{Asset \& Risk Manager} &
	  \textbf{Security Expert} & \textbf{Developer or Administrator} \\
	\midrule
	\citer{cyber} & \textbullet & \textbullet & \textbullet & \textbullet \\
	\citer{ibm_db2} & & & & \textbullet \\
	\citer{making_decisions} & \textbullet & \textbullet & \textbullet & \textbullet \\
	\citer{pmmp} & \textbullet & \textbullet & \textbullet & \textbullet \\
	\citer{complex_path} & \textbullet & \textbullet & & \\
	\citer{handbook} & \textbullet & \textbullet & \textbullet & \textbullet \\
	\citer{ibm_z} & \textbullet & & \textbullet & \textbullet \\
	\bottomrule
  \end{tabular}
\end{table*}

For the \emph{Migration Manager} archetype, the literature mostly agrees on wording and
responsibilities. Other terms in the literature are ``Executive Officer''~\citer{pmmp},
``IT-Manager''~\citer{ibm_z}, and ``Business \& Technology
Manager''~\citer{complex_path}. Regarding skills and responsibilities,
Muller and Heesch~\citer{making_decisions} expect purchasing experience, legal
knowledge, and knowledge about the organization. Attema et al.~\citer{handbook} see
strategy and policy creation as their most important task and von Nethen et
al.~\citer{pmmp} expect them to be project leaders.

The \emph{Asset \& Risk Manager} archetype is also called ``Technology
Executive''~\citer{complex_path}. Muller and Heesch~\citer{making_decisions} and
Attema et al.~\citer{handbook} implicitly define such a role by expecting knowledge
about the organization's data and assets. As further responsibilities of this
archetype, von Nethen et al.~\citer{pmmp} request them to do the risk assessment.
According to Attema et al.~\citer{handbook}, these roles should also be strategy and
policy makers.

The main responsibilities of the \emph{Security Expert} arche\-type are
knowledge about cryptography and the organization's network architecture as well as
its assets~\citer{making_decisions}. Furthermore, these roles should decide how
cryptography is to be migrated~\citer{handbook} and should educate the
organization about PQC and its importance~\citer{pmmp}.

The migration has to be planned meticulously to be successful, but also
executed with precision and competence. For this the \emph{Developer or Administrator}
archetype is especially important as noted by all but one paper.
Additionally, these roles should also have knowledge about the organization's network
architecture and cryptography~\citer{making_decisions}. According to Attema et
al.~\citer{handbook}, they should take part in the decision how cryptography needs to
be migrated.

Every role archetype which is part of the migration team has to be at least informed
about everything happening in each migration step. In most cases, we expect them
to also be consulting the team, thus contributing their expertise. Still, each
archetype will be associated with migration steps for which they are responsible.
Deducing from the literature's definitions of both the migration steps and
roles, we assigned a responsible archetype to each migration step as shown in
Table~\ref{t:role_step}.
\begin{table*}
\small
  \centering
  \caption{The matching of migration phases and the roles responsible for them
		respectively.\label{t:role_step}}
  \begin{tabular}{l!{\color{lightgray}\vrule}cccc}
    \toprule
		\textbf{Step} & \parbox{2.3cm}{\centering\textbf{Migration Manager}} &
			\parbox{2.3cm}{\centering\textbf{Asset \& Risk Manager}} &
			\parbox{2.3cm}{\centering\textbf{Security Expert}} &
			\parbox{2.4cm}{\centering\textbf{Developer /
			Administrator}} \\
	\midrule
	Asset Identification & & \textbullet & & \\
	Risk Assessment & & \textbullet & & \\
	Cryptographic Identification & & \textbullet & & \\
	Cryptographic Assessment & & & \textbullet & \\
	Cryptographic Prioritization & & & \textbullet & \\
	Migration Plan & \textbullet & & & \\
	Execution & & & & \textbullet \\
	Maintenance & & \textbullet & & \textbullet \\
	\bottomrule
  \end{tabular}
\end{table*}

Despite the Migration Manager being responsible for only one migration step, they are
accountable for all of them and are the main role to communicate the team's decisions
and needs to the organization's management.

The Security Expert and the Developer or Administrator archetypes have responsibility for
only two migration steps, as well. Yet, they are both consulting in almost all other
steps. Thereby, the Security Experts' knowledge helps to build a secure solution with
as little security flaws as possible and the Developers' and Administrators'
involvement is necessary to plan a solution that can actually be realized afterwards.

\begin{tcolorbox}[colback=gray!10,colframe=gray!50, title=\textbf{RQ1 Summary},
	coltitle=black]
	We identified the four phases ``Diagnosis'', ``Plan'', ``Execution'', and
	``Maintenance'' as the fundamental phases for migrating software systems towards
	PQC\@.

	We concretized Diagnosis by splitting it up into the identification of assets, risk
	assessment, cryptographic identification, as well as cryptographic assessment. The
	planning phase contains the cryptographic prioritization and the planning for the
	migration.

	We determined that ``Migration Manager'', ``Asset \& Risk Manager'', ``Security Expert'', and
	``Developer/ Administrator'' are the four key role archetypes for the migration and mapped
	the archetypes to the appropriate migration steps. With this approach we unified the
	literature and provide a consensus.
\end{tcolorbox}

\subsection{RQ2: Applications of PQC}
\label{sec:results:rq2}
\begin{table*}
\small
      \renewcommand{\cellalign}{tl}
      \centering
      \caption{Classification of all papers according the application of PQC.\label{t:technological_modifications}}
      \begin{tabular}
        {c!{\color{lightgray}\vrule}
        ccc!{\color{lightgray}\vrule}
        ccccccccc!{\color{lightgray}\vrule}
        cccccccccc}
        \toprule
        \textbf{Ref} &
        \multicolumn{3}{c!{\color{lightgray}\vrule}}{\textbf{Hybrid PQC}} &
        \multicolumn{9}{c!{\color{lightgray}\vrule}}{\textbf{Migrated Software System}} &
        \multicolumn{10}{c}{\textbf{Adjusted Standard}}\\
        \midrule
        & \rotatebox[origin=l]{90}{Claimed} &
          \rotatebox[origin=l]{90}{Recommended} &
          \rotatebox[origin=l]{90}{Mentioned} &
          \rotatebox[origin=l]{90}{Nginx} &
          \rotatebox[origin=l]{90}{IBM DB2} &
          \rotatebox[origin=l]{90}{PostgreSQL} &
          \rotatebox[origin=l]{90}{IoT ecosystem} &
          \rotatebox[origin=l]{90}{ROS} &
          \rotatebox[origin=l]{90}{Hyperledger Fabric} &
          \rotatebox[origin=l]{90}{Delta Chat} &
          \rotatebox[origin=l]{90}{Certification Authority} &
          \rotatebox[origin=l]{90}{Apache Kafka} &
          \rotatebox[origin=l]{90}{IBE} &
          \rotatebox[origin=l]{90}{TLS} &
          \rotatebox[origin=l]{90}{SSH} &
          \rotatebox[origin=l]{90}{S/MIME} &
          \rotatebox[origin=l]{90}{PGP} &
          \rotatebox[origin=l]{90}{X.509} &
          \rotatebox[origin=l]{90}{OCSP} &
          \rotatebox[origin=l]{90}{CMP} &
          \rotatebox[origin=l]{90}{EST} &
          \rotatebox[origin=l]{90}{IPsec}
          \\
          \midrule
        \citer{caraf} &  &  &  &  &  &  &  &  &  &  &  &  &  &  &  &  &  &  &  &  &  & \\
        \citer{cyber} &  &  & \textbullet &  &  &  &  &  &  &  &  &  &  &  &  &  &  & $\circ$ &  &  &  & \\
        \citer{impact_hybrid} & \textbullet &  &  &  &  &  &  &  &  &  & \textbullet &  &  &  &  &  &  & \textbullet & \textbullet & \textbullet & \textbullet & \\
        \citer{microservice} & \textbullet &  &  &  &  &  &  &  &  &  &  & $\times$ &  & $\circ$ &  &  &  &  &  &  & \\
        \citer{tls_database} & \textbullet &  &  &  &  & \textbullet &  &  &  &  &  &  &  & \textbullet &  &  &  &  &  &  &  & \\
        \citer{ibe} &  &  &  &  &  &  &  &  &  &  &  &  & \textbullet &  &  &  &  &  &  &  &  &  \\
        \citer{ibm_db2} &  &  & \textbullet &  & \textbullet &  &  &  &  &  &  &  &  & \textbullet &  &  &  &  &  &  &  &  \\
        \citer{dep_analysis} &  &  &  &  &  &  &  &  &  &  &  &  &  &  &  &  &  &  &  &  &  & \\
        \citer{nist_migration} &  &  & \textbullet &  &  &  &  &  &  &  &  &  &  &  &  &  &  &  &  &  &  & \\
        \citer{making_decisions} &  &  & \textbullet &  &  &  &  &  &  &  &  &  &  &  &  &  &  &  &  &  &  & \\
        \citer{nginx_migration} & \textbullet &  &  & \textbullet &  &  &  &  &  &  &  &  &  &  &  &  &  &  &  &  &  & \\
        \citer{hbs_iot} &  &  &  &  &  &  & $\circ$ &  &  &  &  &  &  &  &  &  &  &  &  &  &  & \\
        \citer{pmmp} &  & \textbullet &  &  &  &  &  &  &  &  &  &  &  &  &  &  &  &  &  &  &  & \\
        \citer{pqc_ros} &  &  &  &  &  &  &  & \textbullet &  &  &  &  &  &  &  &  &  & \textbullet &  &  &  & \textbullet \\
        \citer{pqfabric} & \textbullet &  &  &  &  &  &  &  & \textbullet &  &  &  &  &  &  &  &  & \textbullet &  &  &  & \\
        \citer{pqc_email} &  &  &  &  &  &  &  &  &  & \textbullet &  &  &  &  &  & \textbullet & \textbullet &  &  &  &  & \\
        \citer{cits} &  &  & \textbullet &  &  &  &  &  &  &  &  &  &  &  &  &  &  &  &  &  &  & \\
        \citer{refine_mosca} &  &  & \textbullet &  &  &  & $\circ$  &  &  &  &  &  &  &  &  &  &  &  &  &  &  & \\
        \citer{complex_path} &  & \textbullet &  &  &  &  &  &  &  &  &  &  &  &  &  &  &  &  &  &  &  & \\
        \citer{handbook} &  & \textbullet &  &  &  &  &  &  &  &  &  &  &  & $\circ$ & $\circ$ & $\circ$ & $\circ$ & $\circ$ &  &  &  & $\circ$ \\
        \citer{ibm_z} & \textbullet &  &  &  &  &  &  &  &  &  &  &  &  &  &  &  &  &  &  &  &  & \\
			\midrule
			\midrule
			$\sum~$\textbullet & 6 & 3 & 6 & 1 & 1 & 1 & 0 & 1 & 1 & 1 & 1 & 0 & 1 & 2 & 0
												 & 1 & 1 & 3 & 1 & 1 & 1 & 1 \\
			\midrule
			$\sum~\circ$ & 0 & 0 & 0 & 0 & 0 & 0 & 2 & 0 & 0 & 0 & 0 & 0 & 0 & 2 & 1 & 1 &
										 1 & 2 & 0 & 0 & 0 & 1 \\
			\midrule
		  $\sum \times$ & 0 & 0 & 0 & 0 & 0 & 0 & 0 & 0 & 0 & 0 & 0 & 1 & 0 & 0 & 0 & 0 &
											0 & 0 & 0 & 0 & 0 & 0 \\
      \bottomrule
      \end{tabular}
			\\~\\
      \begin{minipage}[t]{0.7\textwidth}
        \centering
            \textbullet~Criterion applied \qquad
            $\circ$ Criterion described theoretically \qquad
            $\times$ Criterion failed
    \end{minipage}
    \end{table*}

Research question RQ2 deals with the current applications of PQC in the context of our final paper set. This includes the consideration of hybrid migration, the migrated software systems, as well as the adapted standards, protocols, and schemes.

\subsubsection{Hybrid PQC}
\label{sec:results:rq2:hybrid}
After successful migration, PQC is either used in full or hybrid form. Full PQC means
the usage of cryptography that is currently considered quantum-resistant only. The
general consensus for the meaning of hybrid in this context refers to the combination
of quantum-resistant and classical
cryptography~\citer{making_decisions}\cite{hybrid_kex_tls_1_3,ssh_pqc_report}. This
applies both to digital signatures and to key establishment~\citer{making_decisions}.
Attema et al.~\citer{handbook} coins the term ``Hybrid AND'' for this
type of hybrid. Furthermore, they define ``Hybrid OR'' as the use of either
quantum-resistant or classical cryptography, which can ensure backwards
compatibility.

Our review
reflects the extent to which hybrid approaches play a role in the literature. In this
context, we do not distinguish between the two types of hybrids. Instead, we draw a
distinction based on whether the individual articles assert the
implementation of either of these two approaches, advocate for their adoption, or
simply mention the viability of hybrid solutions. This is illustrated in
Table~\ref{t:rq_2}.

For reasons of clarity, an article is only assigned to one of the three
sub-categories. Articles falling under the \emph{claimed} category explicitly
mention hybrid solutions. Six of the 21 articles claim to
provide a hybrid solution by using hybrid certificates~\citer{impact_hybrid},
signature schemes~\citer{microservice,nginx_migration,pqfabric,ibm_z}, or key
establishment~\citer{tls_database,nginx_migration,ibm_z} after the migration. Three
articles advocate for a hybrid solution, while an additional six articles merely
mention it as a potential implementation option. Notably, in the remaining articles,
comprising almost a third of the total, hybrids are not referenced in any manner.

\subsubsection{Migrated Software System}
\label{sec:results:rq2:migratedsoftwaresystem}
We now illustrate the variety of software systems for which a PQC migration was
executed. The systems range from web servers via operating systems all the way
through to databases. It turned out that ten of the 21 articles describe an actual
PQC migration. We therefore created a simple classification (see
Table~\ref{t:technological_modifications}) on whether the migration was either
carried out successfully (criterion applies), described theoretically (criterion
described theoretically) or failed (criterion failed).

Hybrid solutions were implemented for the migration of the Nginx web server, a
certification authority (CA), the Hyperledger Fabric blockchain framework, and the
PostgreSQL database management system, whereas hybrid processes were only mentioned
as a possibility for Db2 but were not discussed further. Holcomb et al.~\citer{pqfabric}
also claim that Hyperledger Fabric's quantum-safe solution
called PQFabric is fully crypto-agile, by which they mean that the used
cryptography can be flexibly replaced by other post-quantum cryptographic
standards.

Weller and van der Gaag~\citer{microservice} attempt to make
Apache Kafka quantum-resistant. However, this was not possible due to integration
issues with the PQC libraries
\emph{Wildfly-OpenSSL}~\footnote{\url{https://github.com/wildfly-security/wildfly-openssl}},
\emph{Bouncy
Castle}~\footnote{\url{https://www.bouncycastle.org/specifications.html}}, and \emph{Open Quantum Safe OpenSSL}~\footnote{\url{https://github.com/open-quantum-safe/oqs-provider}}. For their PQC performance evaluations, the authors relied on the pre-built
PQC client and PQC server solutions of the Open Quantum Safe (OQS) project, which
provides automatically working hybrid solutions. However, we consider the actual
attempted Kafka migration to have failed, which is why we have marked it accordingly
in the table.

\subsubsection{Adjusted Standard}
\label{sec:results:rq2:adjustedstandard}
In addition to the analysis of migration types and migrated software systems,
our examination focuses on the cryptographic standards that were modified
during the migration process. It was ensured that only articles that
describe the adaptation of the standard in detail and in depth are included as
``criterion described theoretically''.

Attema et al.~\citer{handbook} describe in detail why and how TLS, SSH, S/MIME, PGP,
X.509 and IPsec must be adapted to ensure quantum resistance. Fan et
al.~\citer{impact_hybrid} describe the concrete practical implementation for the
migration of a Certificate Authority (CA) in depth using the modifications of X.509,
OCSP, CMP, and EST\@. In addition, they are the only ones that fully address
the necessary technical specifications for the process of issuing, exchanging, and
verifying certificates in the context of the aforementioned protocols and have
successfully implemented the migration. While the ETSI~\citer{cyber} describes the
PQC extension of the X.509 standard in theory, Varma et al.~\citer{pqc_ros} and
Holcomb et al.~\citer{pqfabric} deal with the practical implementation of the
extension of the X.509 certificate. However, these articles do not highlight the
wide-ranging relationships between the various components and protocols compared to the article by Fan et
al.~\citer{impact_hybrid}, which is why they are represented in the table as X.509.

In addition to the Delta Chat messenger, its email infrastructure was
also made quantum-resistant, as messages are also sent via the user's
email~\citer{pqc_email}.
Specifically, the two encryption protocols S/MIME and PGP were successfully adapted
for this purpose. For the actual implementation, rpgp CRYSTALS-Kyber,
CRYSTALS-Dilithium and Picnic were added as public-key encryption and digital
signature schemes. For S/MIME, on the other hand, the \emph{botan}~\footnote{\url{https://github.com/randombit/botan}} library was
extended to include PQC\@.

The Robotic Operating System (ROS) was made quantum-safe by adapting it at both
application and network level~\citer{pqc_ros}. RSA was replaced by Bimodal Lattice
Signature Scheme (BLISS) as the signature scheme, the previous IPsec VPN \texttt{racoon} by the quantum-resistant
\emph{strongSwan}~\footnote{\url{https://github.com/strongswan/strongswan}} IPsec VPN~\citer{pqc_ros}.

For both database management systems PostgreSQL and Db2, PQC was integrated on both
the client and server side. For Db2, CRYSTALS-Kyber was used as key exchange and
CRYSTALS-Dilithium as signature scheme~\citer{ibm_db2}. TLS 1.3 was integrated for
both client and server, which was provided by \emph{IBM Global Security Kit (GSKit)}~\footnote{\url{https://www.ibm.com/docs/en/informix-servers/14.10?topic=protocol-global-security-kit-gskit}}.
For PostgreSQL, the quantum-safe OpenSSL version of Open Quantum Safe was used for the
digital signatures as well as for key exchange~\citer{tls_database}. Furthermore, the
ECDH specifications were hardcoded and had to be adapted.

For identity-based encryption, quantum-secure lattice cryptography (DLPlattice) was
integrated as part of an exemplary REST service. For this purpose, the open-source
C-library \emph{MIRACL}~\footnote{\url{https://github.com/miracl/MIRACL}} as well as custom wrappers and library implementations were used to
ensure the necessary functionality for IBE~\citer{ibe}.
\begin{tcolorbox}[colback=gray!10,colframe=gray!50, title=\textbf{RQ2 Summary},
	coltitle=black]
	PQC\@ solutions, like quantum-resistant key exchanges and digital signature
	schemes, were integrated in ten different software systems, ranging from robotic
	operating systems to web servers or messengers.

	The migration was conducted by exchanging or enhancing classic cryptographic
	standards like TLS, S/MIME or X.509, using either proprietary (e.g., IBM Global
	Security Kit) or open source libraries (e.g., liboqs) that offer implementations of
	algorithms that are currently considered quantum-resistant.

	Over half of the migrated systems use hybrid solutions after the transition,
	making it possible to rely on a combination of classic and post-quantum
	cryptography.
\end{tcolorbox}

\subsection{RQ3: Challenges}
\label{sec:results:rq3}
\begin{table}
  \renewcommand{\cellalign}{tl}
  \centering
  \caption{The challenges according to our definitions and how the literature
		covers them. Papers not listed did not address any migration
	  challenges.\label{t:challenges}}
  \resizebox{1.0\linewidth}{!}{\begin{tabular}
		{c!{\color{lightgray}\vrule}
		ccc!{\color{lightgray}\vrule}
		cccc!{\color{lightgray}\vrule}
		cccc
		}
		\toprule
		\textbf{Ref} &
		\multicolumn{3}{c!{\color{lightgray}\vrule}}{\textbf{Organization}} &
		\multicolumn{4}{c!{\color{lightgray}\vrule}}{\textbf{PQC}} &
		\multicolumn{4}{c}{\parbox{2.4cm}{\centering\textbf{Code and\\Documentation}}}\\
		\midrule &
		\rotatebox[origin=l]{90}{Education \& Expertise} &
		\rotatebox[origin=l]{90}{Time Effort \& Costs} &
		\rotatebox[origin=l]{90}{Sceptics} &
		\rotatebox[origin=l]{90}{Hardware Requirements} &
		\rotatebox[origin=l]{90}{Performance \& Availability} &
		\rotatebox[origin=l]{90}{Security Concerns} &
		\rotatebox[origin=l]{90}{Support \& Features} &
		\rotatebox[origin=l]{90}{Complexity, Variety \& Magnitude} &
		\rotatebox[origin=l]{90}{In Development} &
		\rotatebox[origin=l]{90}{Legacy Encryption \& Environment} &
		\rotatebox[origin=l]{90}{Weakness \& Error-Proneness} \\
		\midrule
	\citer{cyber} &  & \textbullet &  &  &  &  &  &  &  &  & \textbullet \\
	\citer{microservice} &  &  &  &  &  &  & \textbullet &  &  &  &  \\
	\citer{tls_database} &  &  &  & \textbullet &  &  &  &  &  &  &  \\
	\citer{ibm_db2} & \textbullet & \textbullet &  & \textbullet &  &  &  & \textbullet & \textbullet & \textbullet & \textbullet \\
	\citer{dep_analysis} & \textbullet & \textbullet &  & \textbullet &  &  &  & \textbullet & \textbullet & \textbullet &  \\
	\citer{making_decisions} &  & \textbullet &  &  &  & \textbullet &  &  &  &  &  \\
	\citer{nginx_migration} & \textbullet & \textbullet & \textbullet &  & \textbullet &  & \textbullet & \textbullet &  &  &  \\
	\citer{hbs_iot} &  & \textbullet & \textbullet & \textbullet & \textbullet &  & \textbullet &  &  & \textbullet &  \\
	\citer{pmmp} & \textbullet & \textbullet &  &  & \textbullet & \textbullet &  & \textbullet &  &  & \textbullet \\
	\citer{pqfabric} &  &  &  &  &  &  & \textbullet &  &  &  &  \\
	\citer{pqc_email} &  &  &  &  &  &  & \textbullet &  &  &  &  \\
	\citer{complex_path} &  & \textbullet &  &  &  &  &  &  &  &  &  \\
	\citer{handbook} &  & \textbullet &  &  &  & \textbullet &  & \textbullet &  &  &  \\
	\midrule
	\midrule
	$\sum$ & 4 & 9 & 2 & 4 & 3 & 3 & 5 & 5 & 2 & 3 & 3 \\
  \bottomrule
\end{tabular}}
\end{table}

This research question aims to ascertain the challenges associated with the migration
of software systems towards post-quantum cryptography. In contrast to other works
that only superficially deal with the challenges of PQC migration, our selected
articles are based on practical experience and in-depth theoretical descriptions that
specifically address the challenges that arise.

First, we look at the challenges that organizations face as a result of a PQC
migration. Then, we will take a closer look at the challenges that post-quantum
cryptography poses at the current point in time. Finally, we present the challenges that can
arise from an existing code base and documentation of a software system. In particular the
articles~\citer{ibm_db2,dep_analysis,nginx_migration,hbs_iot,pmmp} deal with the
various challenges and accordingly are the source for a large part of the findings.
Table~\ref{t:challenges} represents the challenges as well as their
subcategories.
\subsubsection{Organization}
\label{sec:results:rq3:organization}
We have identified the three subcategories \emph{Education \& Expertise},
\emph{Planning}, and \emph{Sceptics} as representatives for the organizational
challenges.

\emph{Education \& Expertise} refers to the lack of expertise in the PQC environment, as well as the difficulties of providing the best possible education to all affected employees. It is regarded as an essential prerequisite prior to
commencing the realization, as indicated by the articles mentioned
above~\citer{ibm_db2,dep_analysis,nginx_migration,hbs_iot,pmmp}, addressing this
subcategory.

The subcategory \emph{Time, Effort \& Costs} refers to challenging resource planning.
Attema et al.\ and Zhang et al.~\citer{handbook,ibm_db2} emphasize this
observation considering the prevailing uncertainty. In addition, the identification
of stakeholders and the planning of communication also belong to this category. Nine
of the 21 papers address this topic, making it the most frequently mentioned
challenge across all subcategories.

Two articles~\citer{nginx_migration,hbs_iot} describe the \emph{Sceptics'} view of PQC,
with~\citer{nginx_migration} arguing this is due to unknown risk and the associated
high costs. This makes sceptics the least frequently mentioned challenge in the
literature.

\subsubsection{PQC}
\label{sec:results:rq3:pqc}
The literature highlights that Post-Quantum Cryptography (PQC) itself poses various
challenges, with a total of eleven different articles addressing them. Consequently,
we have opted to introduce the PQC category to address these challenges. In our
opinion, these challenges are best examined from the four different perspectives of
\emph{Hardware Requirements}, \emph{Performance \& Availability}, \emph{Security
Concerns}, and \emph{Support \& Features}.

\emph{Hardware requirements} are increasing due to growing computational costs, a
consequence of the expanding dimensions of keys, ciphertext, and signature.
Hasan et al.~\citer{dep_analysis} additionally describe the required replacement of
numerous hardware appliances which integrate hardware acceleration for broadly used
cryptosystems.

The second challenge is reflected in the overall availability of the software systems
and their performance during and after the migration. Von Nethen et al.~\citer{pmmp} emphasize
that systems must remain interoperable to sustain availability across different
systems. Furthermore, Suhail et al.~\citer{hbs_iot} underscore the trade-off in scenarios
with limited resources, balancing the challenges posed by excessive data and the need
for optimal performance.

The \emph{Security Concerns} subcategory reflects the risks that can arise from the use of
current PQC\@. This means either the mathematical procedures on which the current PQC
algorithms are based on can be broken or that the algorithms' implementations contain
vulnerabilities. Muller and Heesch~\citer{making_decisions} as well as Von Nethen et al.~\citer{pmmp}
caution against downgrade attacks, where an attacker tries to divert communication between the sender and
receiver to classical algorithms.

The last sub-category represents the challenges that arise with regard to \emph{support}
of the implemented PQC features. Currently, most libraries that offer PQC are under
development. Some implementations are therefore still incomplete or have faulty
features, such as an incomplete API or a corrupt key store, as described by Weller et al.~\citer{microservice} and Mohammad et al.~\citer{nginx_migration}. 
Missing, insufficient or incorrect documentation and guidelines of current PQC implementations also pose issues. In this context, Suhail et al.~\citer{hbs_iot} emphasize the lack of necessary parameter information for certain PQC schemes, which can complicate or prevent the applicability of PQC.

\subsubsection{Code and Documentation}
\label{sec:results:rq3:codeanddocumentation}
The \emph{Code and Documentation} category includes challenges that may arise from the
current source code and documentation. Within this category, we have identified the
following subcategories: \emph{Complexity, Variety \& Magnitude}, \emph{In Development},
\emph{Legacy Encryption \& Environment}, and \emph{Weakness \& Error-proneness}.

\emph{Complexity, Variety \& Magnitude} illustrate the challenges that can arise due to a
large and complex code base, configuration and infrastructure. Zhang et al.~\citer{ibm_db2}
describe a migration of a code base comprising tens
of millions of lines of C/C++ code. Associated with this is the variety of components
and diverse setup possibilities which in turn also influences the documentation.

The next subcategory \emph{In Development} covers the challenges that can arise due to
ongoing software development. This leads in particular to increasing effort to
maintain an up to date cryptographic inventory~\citer{dep_analysis}.

The next category deals with the challenges that arise due to legacy environment
and the associated legacy encryption. During the data analysis and
synthesis~\ref{s:methodology:data_synthesis}, we saw that maintaining
outdated code bases and infrastructures leads to poorer maintenance, backward
compatibility, and difficult integration of encryption. As a concrete example, the
use of DES in the Db2 environment can be allowed due to the need to support older Db2
clients~\citer{ibm_db2}.

The last subcategory represents the challenges that can arise due to poor code
quality (e.g., hard-coded cryptographic functions or parameters~\citer{ibm_db2}),
missing or inadequate documentation~\citer{ibm_db2}, and the lack of encryption of
assets which require protection~\citer{cyber}.

\begin{tcolorbox}[colback=gray!10,colframe=gray!50, title=\textbf{RQ3 Summary},
	coltitle=black]
	Organizational, post-quantum cryptographic, as well as code- and document-based
	challenges are the three main challenges we identified that emerge while migrating
	towards PQC\@.

	While the first category illustrates challenges in PQC expertise or
	rising time and effort, the second category covers challenges regarding
	security concerns of the currently available PQC implementations as well as missing
	support and features. Code and documentation-based challenges can occur due to the
	complexity and variety of the code base or due to its legacy environment and
	encryption.
\end{tcolorbox}

\section{Discussion}
\label{s:discussion}
In this section we illustrate and discuss the main takeaways of this study, which are:
\begin{enumerate}
    \item Lack of Maturity in PQC implementations;
    \item Lack of an established migration framework;
    \item Lack of concept and specific realization of a cryptographic inventory;
    \item Sparse application of PQC\@;
\end{enumerate}

Before we can delve into the main takeaways, we address the notion of
\emph{crypto-agility}, which is present in the majority of the reviewed papers 
but not in our results. The reason for this deliberate decision is that the
literature has
neither a uniform nor clear definition for crypto-agility. The understanding of
the term may vary from aspects like \emph{Is the software or hardware system
updatable?} or \emph{If RSA-2048 worked for the protocol or implementation, what
does that mean for the use of RSA-4096?} to the far more complex such as \emph{If only
small Diffie-Hellman exchanges were used in the protocol or implementation, can
several hybrid key exchanges be integrated supporting several different key agreement schemes
satisfying all requirements like computational or memory costs?}.

Similarly, literature differs widely in interpretation. For example, White et
al.~\citer{ibm_z} try to further differentiate between the types of
crypto-agility, while others like Mashatan et al.~\citer{complex_path} simply
recommend to invest into it, but do not address the topic in more detail. When it comes
to concrete steps and recommendations to make systems more crypto-agile, the
literature mentions hybrid cryptographic schemes and modular code with little
technical debts, like hard-coded parameters, as desirable~\citer{pmmp}. All these
factors are already addressed in great detail in Section~\ref{s:results}.

One secondary study~\cite{on_ca} tried to find a common definition of crypto-agility.
Their conclusion is that a system or infrastructure is crypto-agile if it supports to
be migrated with comparatively low effort. 
As a very detailed and concrete defintion for crypto-agility is hard to grasp at the time, we agree to focus on the basic question of an alleviated transition and therefore see the question of introducing more complex crypto-agile mechanisms as being redundant for now.
Therefore, we focus on how to ease migration and discuss our
main takeaways in the following.

\subsection{Lack of Maturity in PQC Implementations}
Within the scope of RQ2 (see Section~\ref{sec:results:rq2}) and RQ3 (see
Section~\ref{sec:results:rq3}), we identified the current PQC implementations'
insufficient maturity and a lack of detail in the underlying standards. Both
aspects negatively impact real-world realizations built upon them.

One of the well-established organizations is the \emph{National Institute of
Standards and Technology (NIST)}\footnote{\url{https://www.nist.gov}}, which
initiated a process for evaluating and standardizing post-quantum cryptographic
algorithms to provide foundational security guarantees. Despite comprehensive
scrutiny, Attema et al.~\citer{handbook} highlight that vulnerabilities are
discovered frequently in the proposed algorithms.

Besides the failings of the proposed standards, the present implementations of PQC algorithms pose a major
challenge as they exhibit deficiencies such as missing or insufficient support for
different programming languages and frameworks (see
Section~\ref{sec:results:rq3:pqc}). Weller et al.~\citer{microservice}, for example,
did not manage to integrate the PQC libraries \emph{Wildfly-OpenSSL} and \emph{Bouncy
Castle} into their microservice environment due to missing and faulty 
functionalities in these libraries.

A large number of PQC libraries are currently under
development and are being actively maintained, such as
\emph{liboqs}~\footnote{\url{https://github.com/open-quantum-safe/liboqs}} from the
\emph{Open Quantum Safe (OQS) project},
\emph{Botan}, and
\emph{PQClean}~\footnote{\url{https://github.com/PQClean/PQClean}}. Which quality
they can achieve and which of them will be adopted most widely remains to be seen.

Another key consideration in this context is the provisioning of PQC libraries for
embedded devices, covered in Section~\ref{sec:results:rq3:pqc}.
Ott et al.~\cite{migration_challenges_ott} also highlight this as a key
challenge due to their constrained computational power and memory size. According to
Barker et al.~\cite{getting_rdy_for_pqc_barker}, this could be possibly counteracted
if the \emph{NIST} standards take hardware constraints into account.
In contrast, Ott et al.~\cite{migration_challenges_ott} criticise the
\emph{NIST} PQC submissions as not ready to use in the real-world.

To conclude this point, on the one hand it is clear that a lot of effort being put into
the matter by experts in terms of theoretical as well as practical work and a 
wide variety of complex problems especially regarding implementation is tackled step by step.
On the other hand, we see in our analysis just how difficult the field of applied cryptography
is and that there is no consensus on how to approach the implementation of PQC algorithms,
in particular for resource-constrained systems. We still believe that these efforts will converge into reliable and
field-tested solutions and implementations, but due to the current lack of practical
experience, most available implementations have to be considered experimental at this point in time.

\begin{tcolorbox}[colback=gray!10,colframe=gray!50, title=\textbf{Key Takeaway},
	coltitle=black]
    Overall, further standardization efforts as well as mature implementations with full
    support for different hardware, programming languages, and frameworks are necessary
    to ensure an efficient realization of PQC.
\end{tcolorbox}

\subsection{Lack of an Established Migration Framework}
Our second main takeaway is the absence of an established migration framework, which
is illustrated by the results of Section~\ref{sec:results:rq1}.

Firstly, the dissimilar terminology for the various migration steps presents a
challenge. Secondly, the activities that characterize these steps are defined in
highly divergent ways.

According to Cryptovision~\cite{eviden_pqc_migration_guide}, the migration
procedure comprises six different phases, namely the project setup, the creation of a
cryptographic inventory, risk understanding, impact assessment, the involvement of
executives, and the execution of the migration. Contrary to this, Alnahawi et
al.~\cite{on_the_state_of_pqc_migration_alnahawi} distinguish
between algorithm migration and system migration. They briefly mention the relevance
of appropriate planning, prioritization, and the creation of a cryptographic
inventory in the context of system migration. This dissent is representative for the
diversity we saw when analyzing our final paper set for the existing
migration procedures. It underlines the importance of a unified migration
process with precisely defined steps. To counteract this lack of consensus, we
identified four migration phases and several sub-steps as fundamental based
on the entirety of our final paper set (see Section~\ref{sec:results:rq1:stepsterminology}).

We further observe that guidelines and recommendations for PQC
migrations are usually brief (see, e.g., \cite{bsi_migration_recommendations,migration_to_pqc_cisa}). 
These documents help the migration team with the first
steps and give some direction but fall short of a well-defined, fleshed-out migration
process.

In our opinion, those are major issues, which lead to practical PQC migrations often
following a trial-and-error approach rather than a clearly defined procedure. We have
noticed this in several studies, such as those by Holcomb et al.~\citer{pqfabric},
Varma et al.~\citer{pqc_ros}, Mohammad et al.~\citer{nginx_migration}, or Weller et
al.~\citer{microservice}.

In addition to the individual activities, a description of the PQC migration process
should include defined roles to ensure that the right stakeholders are involved. 
We condensed the role definitions of the theoretical frameworks in our
surveyed literature into a clear and comprehensive role structure. Additionally, we
mapped this role concept to the previously extracted and canonized role framework in
Section~\ref{sec:results:rq1:roles}. Still, these role assignments lack verification
from real-world experience. None of the performed migrations described by our final
paper set mentioned any tangible role allocation. Instead, they were mainly performed by
the authors, without addressing roles.

To conclude, current migration frameworks mostly provide highly abstract
descriptions. Quite often hints about the actual steps are given, e.g., by
referring to some tasks and tools that might be handy in building a
cryptographic inventory. Yet, a hands-on or step-by-step guide is missing and
left to (the individual needs of) the reader. It is likely that these
frameworks will become more concrete as more and more entities approach the
migration and share their experiences. Right now, even the more detailed
frameworks leave a lot of room for interpretation.

\begin{tcolorbox}[colback=gray!10,colframe=gray!50, title=\textbf{Key Takeaway},
	coltitle=black]
		A consensus on the various migration steps and
		roles involved for successfully conducting a PQC migration is missing. We
		summarized the current status-quo of theory and practice in four steps and
		assigned them appropriate roles (see
		Section~\ref{sec:results:rq1:literaturesteps} and
		Section~\ref{sec:results:rq1:roles}). Multiple structured and comprehensive
		migrations of software systems which follow our specified concept are necessary
		for its validation and refinement to shape an established migration framework.
\end{tcolorbox}

\subsection{Lack of Concept for and Specific Realization of a Cryptographic Inventory}
The cryptographic inventory is an essential artifact which contains the necessary
information about the cryptography used for various assets. As described in
Section~\ref{sec:results:rq1:literaturesteps}, most of the literature mentions the
compilation of the cryptographic inventory. In particular Hasan et
al.~\citer{dep_analysis}, Barker et al.~\citer{nist_migration}, Attema et
al.~\citer{handbook}, and ETSI~\citer{cyber} describe the cryptographic
inventory and its content in more detail.

There is a no consensus on what constitutes a cryptographic
inventory and how it differs from other artifacts, such as an asset inventory. The
inconsistent use of terminology in the literature which sometimes refers to
processes and sometimes to artifacts, adds to the confusion and makes it difficult to
develop a standardized understanding of this concept.

Beyond our final paper set, this is also confirmed by the Cryptosense
SA~\cite{crypto_inventory_cryptosense} and FS-ISAC
Inc.~\cite{infrastructure_inventory}. The former see the cryptographic inventory as
the combination of the infrastructure and applied
cryptography~\cite{crypto_inventory_cryptosense}, whereas the latter simply mention
the need to examine all existing inventories regarding
cryptography~\cite{infrastructure_inventory}.

We therefore provide a consistent terminology for the various process steps (see
Section~\ref{sec:results:rq1:stepsterminology}) by distinguishing between asset
identification as well as cryptographic identification, evaluation, and
prioritization based on the literature we reviewed.

So far, the cryptographic inventory has only been described at a conceptual and
theoretical level. There is a lack of practical guidance or recommendations as to
which tools are most suitable, to what extent, and under which conditions.

An initial approach for tackling this issue is provided by
Cryptosense~\cite{crypto_inventory_cryptosense}. For the determination of the used
cryptography (e.g., rest encryption, file verification, or password protection), they
recommend performing manual scanning by reviewing code or interviewing developers,
static scanning of the source or byte code, as well as run-time tracing to observe
the application behaviour during execution. They propose the tool ``Cryptosense
Analyzer'' for the static and dynamic application security testing. To determine an
infrastructure's cryptography, they recommend examining the network traffic for the
used protocols, the file system for certificates and keys, as well as the hardware
security modules which often contain master keys of the organization. For capturing
the cryptography of the network traffic, the authors suggest the tools
``testssl.sh''~\footnote{\url{https://github.com/drwetter/testssl.sh}} for TLS,
``ssh\_scan''~\footnote{\url{https://github.com/mozilla/ssh_scan}} for SSH, and
``ike-scan''~\footnote{\url{https://github.com/royhills/ike-scan}} for IPsec.

To conclude, we expect the step of building a cryptographic inventory to be
split into various subtasks including steps like surveys and the use of questionnaires
e.g., for system administrators, network scanning in various ways, different
approaches for code analysis and many more. We hope for the creation of
common standards for building inventories, e.g., for a suitable data format.
Similar problem statements and approaches may be found in fields like the
modern \emph{Software Bill of Materials (SBOM)} (compare e.g., the US Executive
Order on Improving the Nation's Cybersecurity~\cite{white_house_executive}) or the even more specific
\emph{Cryptography Bill of Materials (CBOM)}\footnote{\url{https://github.com/IBM/CBOM}}.

\begin{tcolorbox}[colback=gray!10,colframe=gray!50, title=\textbf{Key Takeaway},
	coltitle=black]
	The concept of \emph{cryptographic inventory} is currently ill-defined and used
	with very different meaning in the literature. What is needed are an in-depth
	examination of the concept, clear definitions, a unified
    terminology, as well as a practice-oriented design. Emerging tooling as
	well as related areas indicate that this area is starting to receive the
	necessary attention.
\end{tcolorbox}

\subsection{Sparse Adaption of PQC}
As we have shown in Section~\ref{sec:results:rq2}, the migration to PQC in software
systems has been extremely rare. Only eleven papers mentioned concrete applications,
while ten others focused on the theoretical description of the migration.

In Section~\ref{sec:results:rq2:hybrid}, we highlighted that there are few
hybrid approaches in which PQC is used in combination with classic cryptography
methods. Since such hybrid approaches are recommended and mentioned so rarely,
they are only used infrequently in practice.

Nonetheless, a lot of preparation has been done lately. Standardization bodies are
actively working on quantum-resistant solutions. For example, many working
groups within the IETF and IRTF added PQC to their work and charters
(e.g., the \emph{Crypto Forum Research Group (CFRG)} which published
RFCs for hash-based signatures~\cite{rfc8391,rfc8554} or \emph{Limited Additional
Mechanisms for PKIX and SMIME (LAMPS)} which made PQC an explicit charter
point\footnote{\url{https://datatracker.ietf.org/wg/lamps/about/}}) and even an
explicit post-quantum working group \emph{Post-Quantum Use in Protocols
(pquip)}\footnote{\url{https://datatracker.ietf.org/wg/pquip/about/}} was founded.
Also, first PQC-related protocol extensions have been published, e.g., by the
\emph{IP Security Maintenance and Extensions~(ipsec\-me)} group~\cite{rfc9242,rfc9370}.
Yet, they have also met significant criticism for their approaches~\cite{pq_ike} and
first implementations are sparse and sometimes do not necessarily comply to
the full extend or all details of the
standard\footnote{\url{https://datatracker.ietf.org/meeting/111/materials/slides-111-ipsecme-combined-slideset-01}}
(compare, e.g., strongSwan).

NIST expects to publish the first post-quantum FIPS
standards\footnote{\url{https://www.nist.gov/news-events/news/2023/08/nist-standardize-encryption-algorithms-can-resist-attack-quantum-computers}}
in the course of 2024. The only encryption scheme that will be
standardised, recommended, and required by regulative means is Kyber aka ML-KEM\@.
Other agencies, such as the German BSI, announced that despite having different
recommendations as of now, they will likely adopt at least some aspects of
the coming NIST standards like approving Kyber aka ML-KEM (for NIST security
levels 3 and 5)~\cite{bsi_tr021021}. We thus expect this to be the starting point
of a wider adoption in practice, as many initial solutions will then focus on
NIST-approved algorithms as long as they are practical enough for real-world use.
This can be seen already, e.g., regarding drafts proposed to the IETF
(compare Julien et al.~\cite{ietf-lamps-cms-kyber-03}, Schwabe et
al.\cite{cfrg-schwabe-kyber-04}, Turner et
al.\cite{ietf-lamps-kyber-certificates-03}, Westerbaan et
al.\cite{tls-westerbaan-xyber768d00-03} to name just a few). 

To conclude, we see more
and more PQC standards being published or heading for publications. We hope for a
kind of domino or snowball effect once the first solutions found their way into
practical use. The additional experience will help to improve the post-quantum
solutions and lead to even further spread of use.

\begin{tcolorbox}[colback=gray!10,colframe=gray!50, title=\textbf{Key Takeaway},
	coltitle=black]
    Few standards regarding PQC exist with many on the way. For the ones that
	already exist, practical adoption is sparse, e.g., due to the
	implementational complexity needed for extending existing standards and
	implementations. A faster development is expected, once regulative
	authorities like NIST release first standards and recommendations.
\end{tcolorbox}

\section{Threats to Validity}

We follow Ampatzoglou et al.~\cite{ampatzoglou2020guidelines} in categorizing the relevant threats to validity of our study. They split the threats to validity for a secondary study such as our SLR into three categories: \emph{study selection validity}, \emph{data validity}, and \emph{research validity}. Most of the threats listed by Ampatzoglou et al.\ are addressed in our \emph{quality assurance plan}, shown in Table~\ref{tab:quality_assurance_plan_table}. In addition, we followed the PRISMA~2020 statement~\cite{prisma_2020} to ensure that we used current state-of-the-art methods throughout our study.

\paragraph{Study Selection Validity}
The process illustrated in Section~\ref{s:methodology} addresses most of the threats to validity Ampatzoglou et al.\ list in this category. In particular, we have used several mitigation mechanisms to make sure we identify relevant studies, including mixing digital libraries as well as venues, using pilot searches and snowballing. We also consider how we have applied inclusion and exclusion criteria as rigorous since we performed random screening and comparison. We have not considered sources not written in English, but do not consider this a shortcoming since we only found very few resources in other languages.

\paragraph{Data Validity}
Our final paper set is relatively small with 21 papers, but also highly relevant for our research questions. It covers a broad set of venues as well as grey and academic literature. We therefore consider it representative and sufficient to answer our research questions. While we did not perform quality assessment of the articles, our inclusion and exclusion criteria were designed to only include papers that have a research character and provide concrete methodologies, frameworks, etc., thus excluding many lower-quality publications such as marketing material. As detailed in our quality assurance plan, data extraction has been performed by several researchers and we have performed random quality checks to ensure unbiased data extraction.

\paragraph{Research Quality}
We have documented our research process extensively and make it available in our supplementary material. This includes the research protocol that we have followed and reported in Section~\ref{s:methodology}. We believe the chosen research method is suitable for answering our research questions since it was our explicit and stated goal throughout to understand the state-of-the-art and the state-of-the-practice as reported in the literature. We use related work throughout this report, but in particular in Section~\ref{s:discussion} where we discuss our findings in relation to the related work. We also believe our study to be generalizable as we capture a broad understanding of the current literature.

\section{Conclusion}
In the face of potential threats posed by quantum computers the urgent need for
quantum-safe cryptographic solutions is evident. Despite global efforts to develop
such solutions, migrating the extensive and diverse software systems of today's
digital infrastructures towards post-quantum cryptography remains a substantial
challenge. We, therefore, conducted a systematic literature review according to the
well-established PRISMA 2020 statement~\cite{prisma_2020}. 

The contribution of this paper is a comprehensive over\-view and entry point into 
the complex topic of PQC migration. We provide insights into the current state-of-the-art
and state-of-the-practice and distill knowledge from a number of academic and grey
sources. In particular, we provide the first generic description of a post-quantum
migration process (RQ1), an overview of which applications have already been migrated
to PQC (RQ2), and investigate the challenges that arise during the migration process (RQ3).

In RQ1 we identified ``Diagnosis'', ``Plan'', ``Execution'', and ``Maintenance'' as
the four fundamental phases comprising more concrete steps for migration based on the
literature. Additionally, we delineated four essential key roles for migration and
mapped each role to its corresponding step in the migration process.

In RQ2, we illustrated ten already migrated software systems and highlighted the
impact of hybrid solutions in this context. These migrations involved the
substitution or enhancement of traditional cryptographic standards, utilizing either
proprietary or open-source libraries.

We furthermore determined that organizational, post-quantum cryptographic, as well as
code- and document-based challenges are fundamental in RQ3 and explained their nature
and origins. Thus, we help understanding and identifying those risk factors while
planning or conducting an actual PQC migration which allows for proper mitigation.

As described in depth in  Section~\ref{s:discussion}, the current lack of maturity in
PQC implementations, the lack of an established migration framework, and a missing
concept for the concrete realization of a cryptographic inventory contribute
to the sparse adaption of PQC in software systems. These obstacles must be
counteracted as quickly as possible to ensure secure PQC solutions that can be
implemented broadly.

\bibliographystyler{IEEEtran}
\bibliographyr{bibliography}
\bibliographystyle{IEEEtran}
\bibliography{bibliography}

\newpage
~\newpage

\begin{IEEEbiography}[{\includegraphics[width=1in,height=1.25in,clip,keepaspectratio]{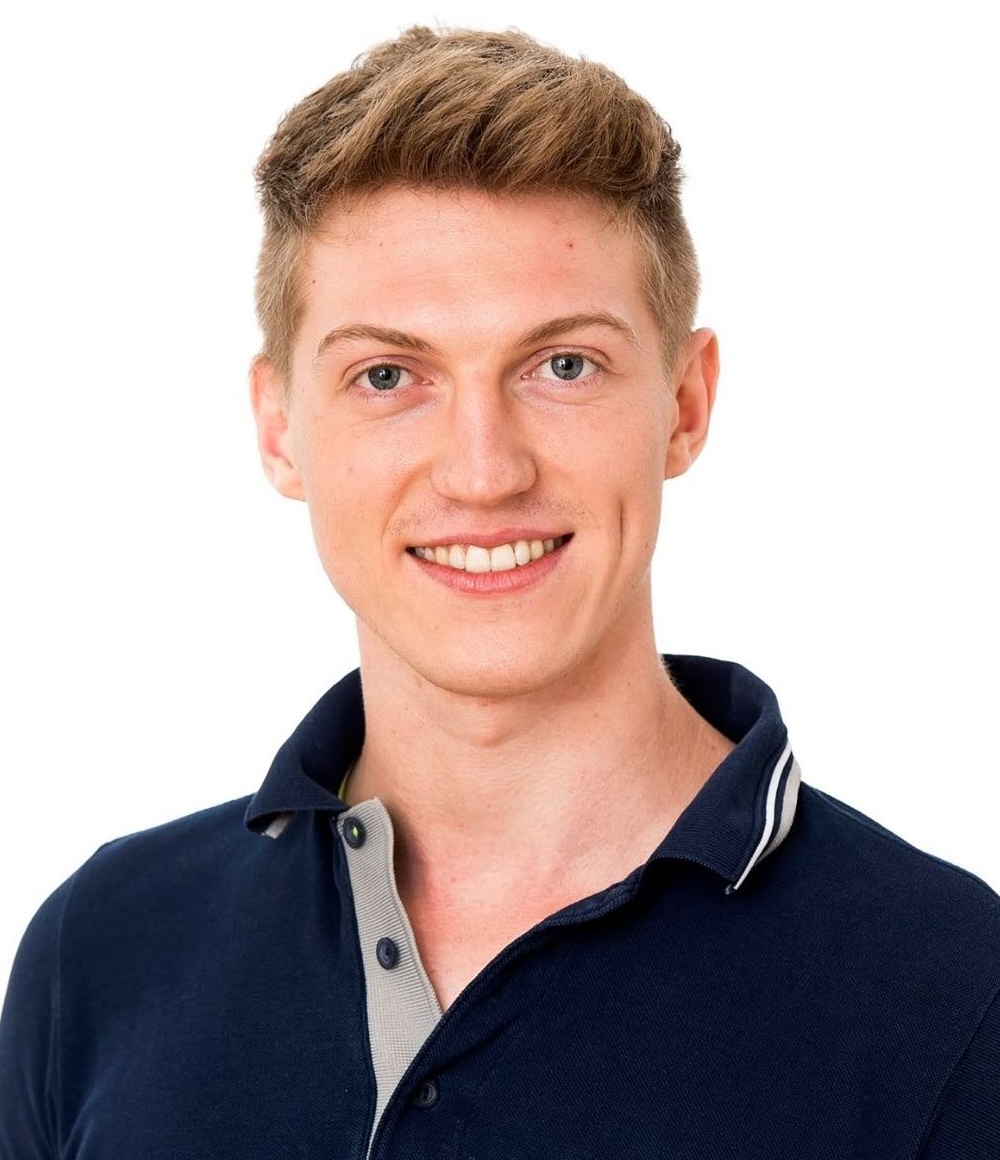}}]
  {Christian Näther} received his Master's degree in Computer Science from the University of Augsburg, Germany.
  As a former software engineer he has gained several years of experience in the fields of software development, distributed systems as well as information security. 
  He is currently a security researcher at XITASO and part of the research project AMiQuaSy, which focuses on the migration of systems towards post-quantum cryptography.
  Christian Näther is an active board member of XITASO’s information security community, where he promotes both internal and customer-specific security. 
\end{IEEEbiography}
  
\begin{IEEEbiography}[{\includegraphics[width=1in,height=1.25in,clip,keepaspectratio]{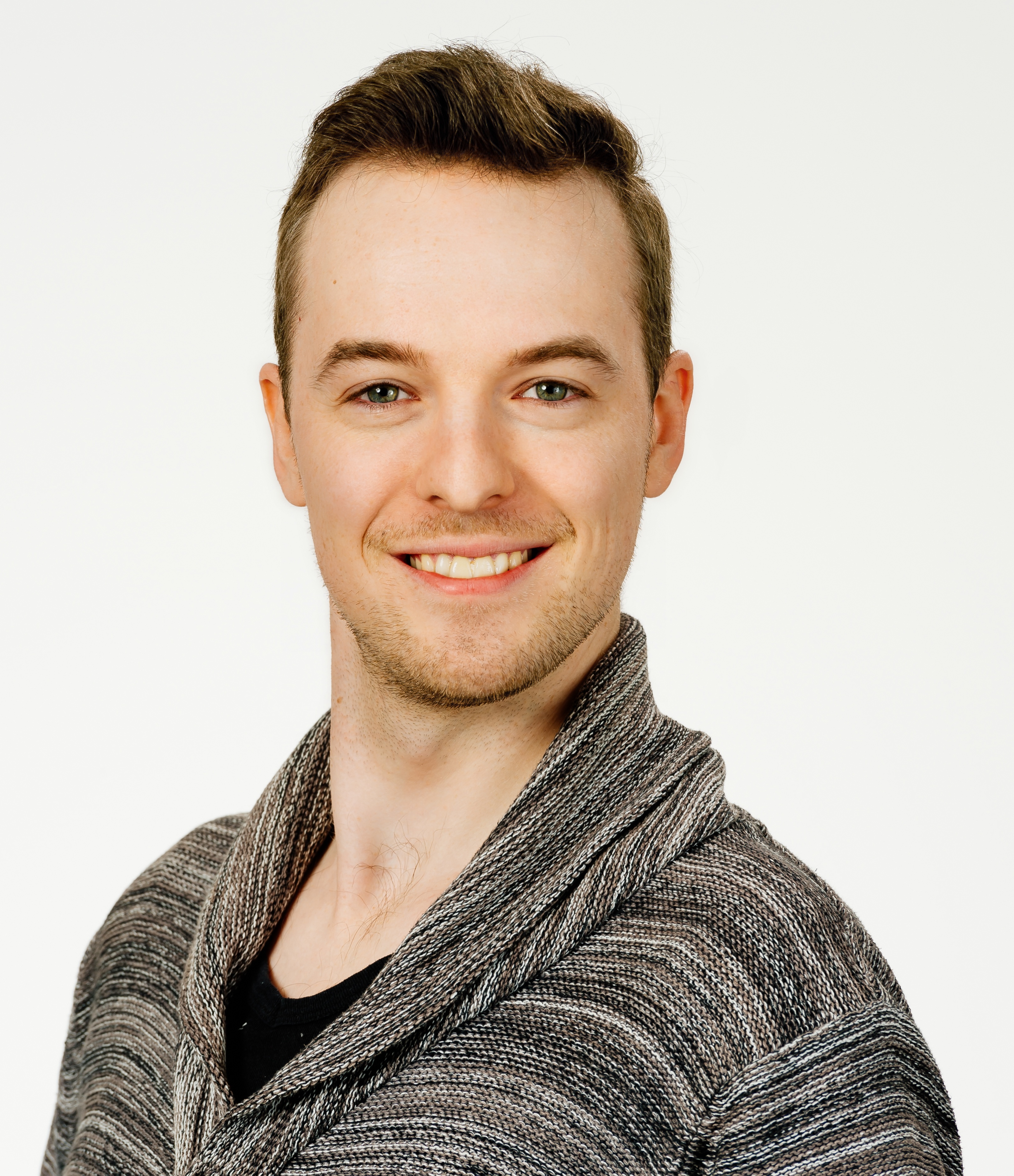}}]
{Daniel Herzinger} holds a Master of Science in Computer Science.
He started working on post-quantum cryptography (PQC) and their adoption into real-world scenarios during his Master's thesis in 2020, together with the German IT security company genua, where he started his IT security journey in 2015. After finishing his studies, he first dived into supporting customers with securing their IT and OT infrastructures at genua's consulting department for two years. He then indulged in his passion for PQC by joining the research group in 2023, focusing on the transition of complex systems and infrastructures to quantum resistance.
\end{IEEEbiography}

\begin{IEEEbiography}[{\includegraphics[width=1in,height=1.25in,clip,keepaspectratio]{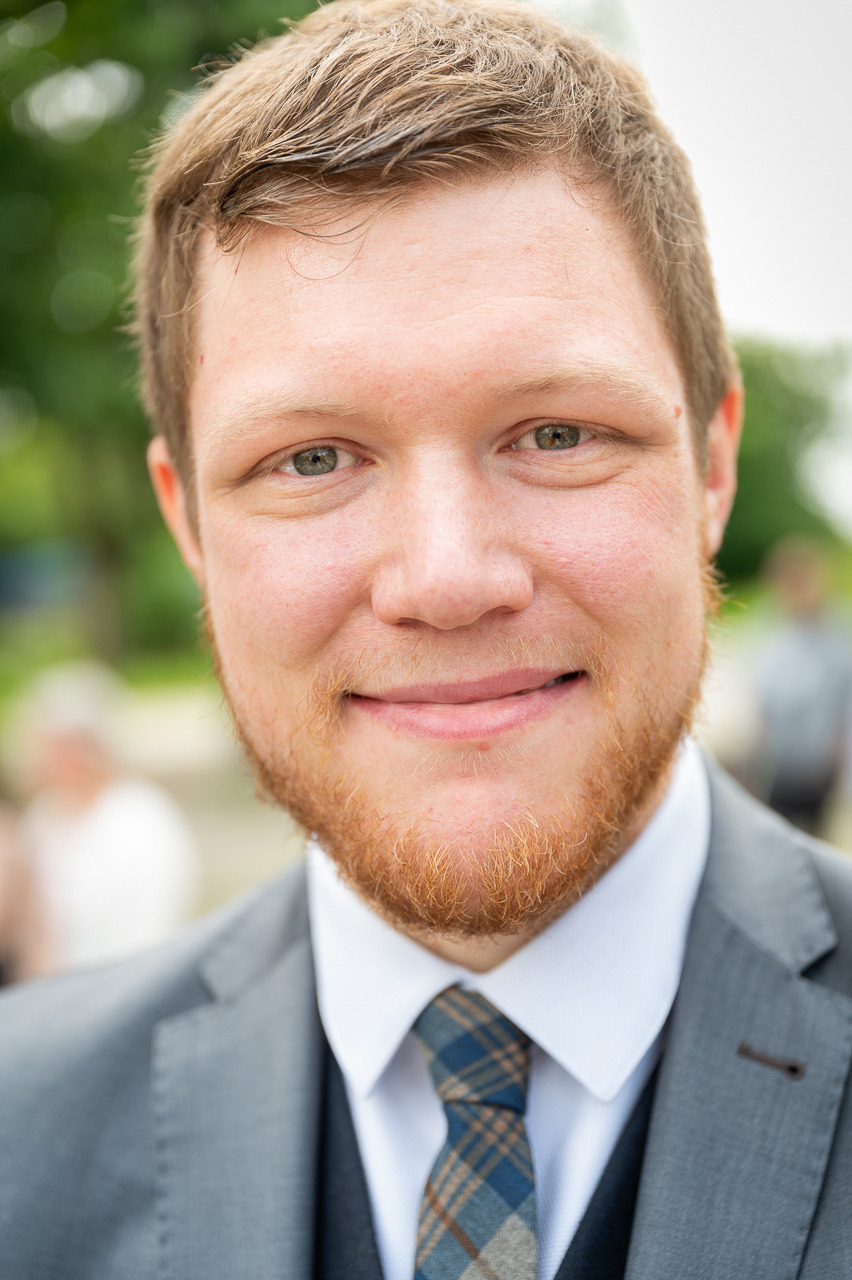}}]
{Stefan-Lukas Gazdag} holds a Master of Science in Computer Science. He focuses on post-quantum cryptography (PQC) and future-proof security mechanisms since 2013, overcoming obstacles in practical use. Being part of the German IT security company genua’s research group he currently works on the third publicly funded PQC research project and other collaborations, enabling the transition to quantum-safe networks. He was e.g. involved in the publication of the first explicit PQC RFC, RFC 8391 describing the signature scheme XMSS, recommended e.g. by US NIST, NSA and German BSI.
\end{IEEEbiography}

\begin{IEEEbiography}[{\includegraphics[width=1in,height=1.25in,clip,keepaspectratio]{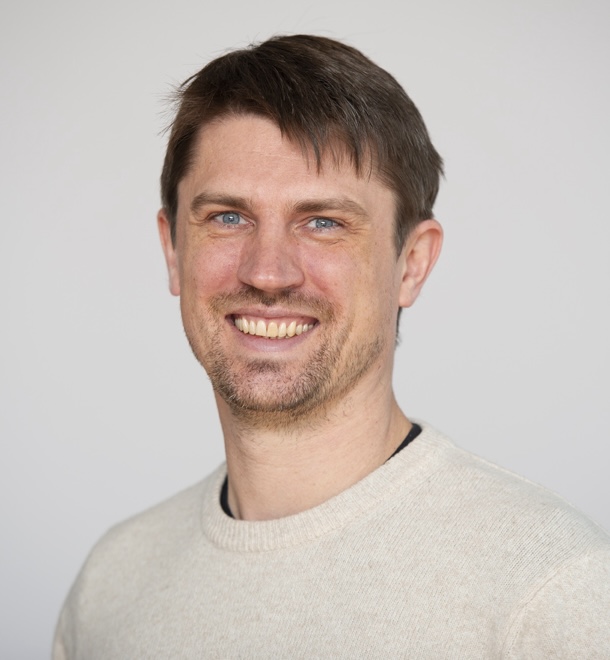}}]
{Jan-Philipp Steghöfer} (Member, IEEE) received the Ph.D. degree in Computer Science from the University of Augsburg, Germany.
As a former professor at the University of Gothenburg he has an extensive and distinguished academic background with a strong passion for research and innovation.
His research interests include security assurance, security of AI systems, agile development in regulated domains, and model-driven engineering.
He is currently a Senior Researcher at XITASO as well as an active member of the software engineering community where he is part of the organisation committee for events like the IEEE International Requirements Engineering Conference (RE). 
He also reviews for International Conference on Software Engineering (ICSE), IEEE Transactions on Software Engineering (TSE), Journal of Systems and Software (JSS), Empirical Software Engineering and Measurement (ESEM), and many other venues.
\end{IEEEbiography}

\begin{IEEEbiography}[{\includegraphics[width=1in,height=1.25in,clip,keepaspectratio]{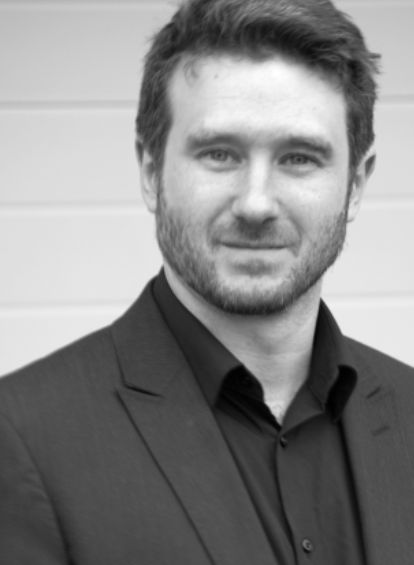}}]
{Simon Daum} completed his doctorate at the Chair of Computational Modeling and Simulation at the Technical University of Munich in 2018 and continued his research in this field until 2021. Since then, he has been head of research at genua GmbH and is responsible for the strategic selection and successful implementation of a wide range of research topics in the field of IT security.
\end{IEEEbiography}

\begin{IEEEbiography}[{\includegraphics[width=1in,height=1.25in,clip,keepaspectratio]{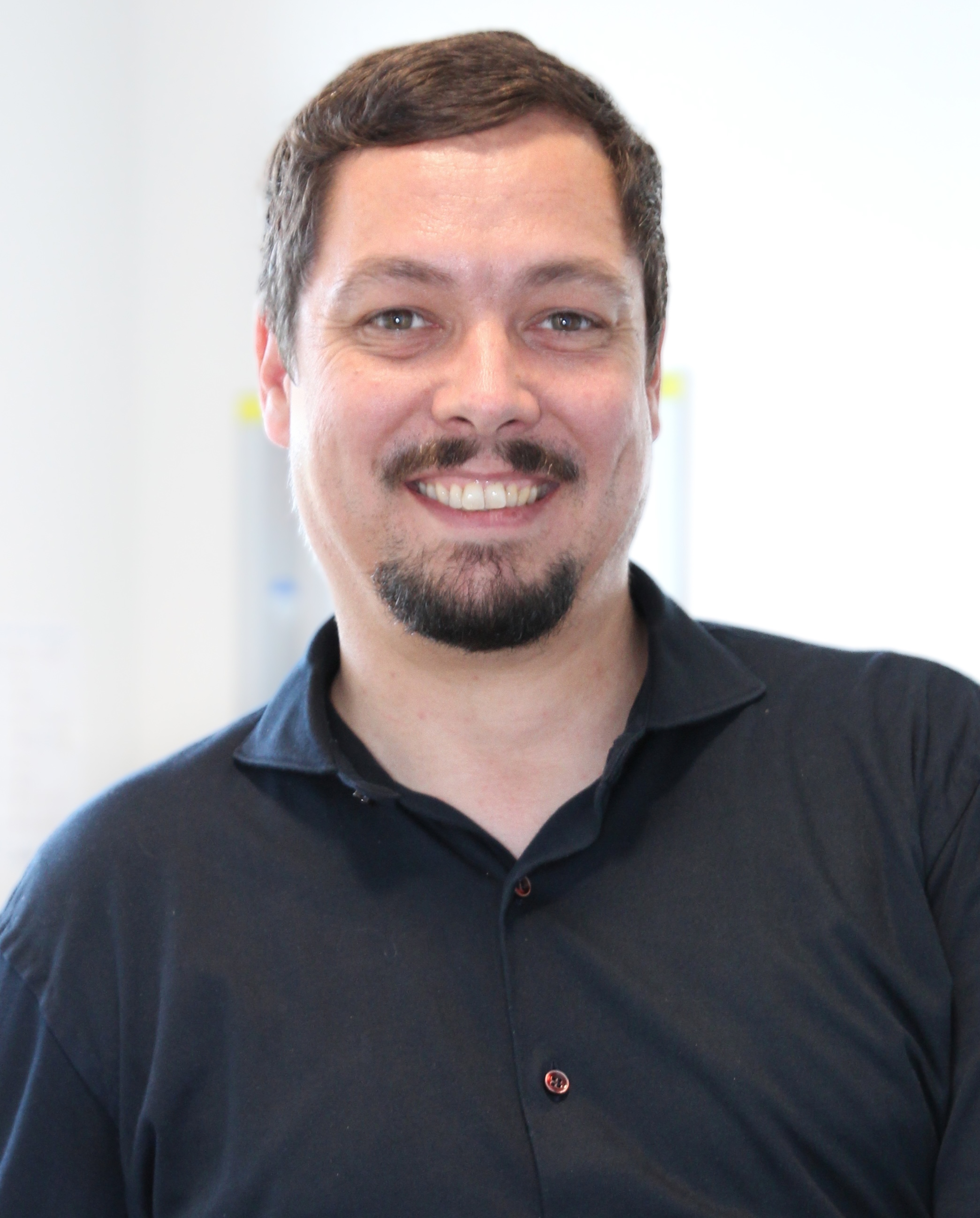}}]
{Daniel Loebenberger} received his doctorate in cryptography from the University of Bonn in 2012, where he conducted research and taught until the end of 2015. From 2016 to 2019, he worked as an IT security expert with a focus on cryptography at genua GmbH, a subsidiary of the Bundesdruckerei GmbH, the German Federal Printing Office, on various topics of professional high-security components. Since January 2019, Daniel Loebenberger has been appointed Professor of Cybersecurity at the Technical University Amberg-Weiden and also heads the department "Secure Infrastructure" at Fraunhofer AISEC's Weiden site. In the lab there, topics of applied post-quantum cryptography and quantum-safe infrastructures are addressed in research and teaching.
\end{IEEEbiography}

\EOD

\end{document}